\documentclass[fleqn,useAMS,usenatbib]{mn2e}

\usepackage{amsmath}
\usepackage{graphicx, amssymb}
\usepackage{times}

\pdfoutput=1

\newcommand{\beq}{\begin{equation}}
\newcommand{\eeq}{\end{equation}}
\newcommand{\bea}{\begin{eqnarray}}
\newcommand{\eea}{\end{eqnarray}}

\title[Dwarf Spheroidal Formation after Reionization]{ Dwarf Spheroidal Satellite Formation in a Reionized Local Group}

\author[Milosavljevi\'c \& Bromm]{Milo\v s Milosavljevi\'c and Volker Bromm \\ Department of Astronomy, The University of Texas at Austin,
  2515 Speedway, Stop C1400, Austin, TX 78712 }

\begin{document}

\label{firstpage}

\maketitle
\topmargin-1cm

\begin{abstract}

Dwarf spheroidal satellite galaxies have emerged 
a powerful probe of small-scale dark matter clustering and of cosmic
reionization. They exhibit structural and
chemical continuity with dwarf
irregular galaxies in the field and with spheroidal galaxies in high-density environments.  
By combining empirical constraints 
derived for star
formation at low gas column densities and metallicities in the
local universe with a model for dark matter and baryonic mass
assembly, we provide an analytical description of how the dwarf
spheroidals acquired their stellar content.  Their progenitors 
formed stars until the gas content, initially reduced from the
cosmic average by the thermal pressure of the reionized intergalactic medium,
was finally ram pressure stripped during the progenitors' accretion on
to
the host galaxy.  Dwarf spheroidal
satellites of differing luminosities seem to share very similar most massive progenitor
histories that reach thresholds for gas cooling by atomic line
emission at epochs at which the Lagrangian volume of the Local Group
should have been reionized. We hypothesize that dwarf spheroidals
formed the bulk of their stars in partially rotationally supported
H\,I disks in a reionized universe.  This model provides an
explanation for the
``common mass scale'' relation and
reproduces the empirical luminosity-size and luminosity-metallicity
relations. Explosive feedback phenomena, such as outflows
driven by the concerted action of supernovae, need not have been
significant in the dwarf spheroidals' formation.  We further speculate
that the true pre-reionization fossils should exhibit a structure distinct from that of
the dwarf spheroidals, e.g., in the form of dense isolated or nuclear
star clusters.

\end{abstract}


\section{Introduction}

The faint satellite galaxies around the
Milky Way and 
Andromeda, the dwarf spheroidals, have been recognized as windows into the clustering of
dark matter on small spatial scales
\citep[e.g.,][]{Lin:83,Zentner:03,Gilmore:07,BoylanKolchin:12}. They
can also be seen as opportunities for learning about the physics of star formation in the
regime in which it is the least efficient, at low gas column densities
and low metallicities, where the star formation is the most susceptible to
environmental influences such as those arising from cosmic reionization.  The discovery of the ultra-faint
dwarf spheroidal satellites around the Milky Way \citep[e.g.,][]{Willman:05a,Willman:05b,Belokurov:06b,Belokurov:07,Sakamoto:06,Zucker:06,Irwin:07,Walsh:07} has pushed our ability
to measure dark matter mass down to $\sim10^6\,M_\odot$ on spatial
scales of only a few tens of parsecs \citep[e.g.,][]{Walker:09}, and has provided us with
examples of the smallest and oldest stellar systems known to possess
their own dark
matter halos \citep[e.g.,][]{Brown:12}.  Here, in an attempt to
explain the structural and chemical
trends seen in the dwarf spheroidal population, we present an
analytical model combining constraints derived from 
investigations of star formation in the local universe with an
idealized 
treatment of the objects' dark and baryonic mass assembly
histories in the standard $\Lambda$ cold dark matter ($\Lambda$CDM) cosmology.

While the dwarf spheroidal
satellite galaxies are dark matter dominated, they
are arguably the most primitive stellar systems in
the local universe, in terms of the relatively small number of stellar generations
that entered their formation and the ancient
origin of at least some of these stars.  The ancient origin, which follows from
color-magnitude relation analysis, and indirectly, from the abundances of
nucleosynthetic tracers, presents an opportunity for
learning about how reionization influenced
star formation in the Local
Group.  The epoch and progress of reionization in the Local
Group and in the universe overall are poorly constrained,
but it is expected that the dark matter halos hosting dwarf
spheroidals had shallow enough gravitational potential wells for
reionization to have reduced
their gaseous baryonic content \citep[e.g.,][]{Bullock:00}.  What fraction of the dwarf
spheroidals' stars formed before reionization has been the subject
of much investigation, often with coarse-grained cosmological collisionless
$N$-body or hydrodynamic simulations aided by subgrid prescriptions specifying the
rate of star formation and the intensity of the resulting feedback \citep[e.g.,][]{Gnedin:06,Bovill:09,Bovill:11a,Bovill:11b,Munoz:09,Busha:10,Li:10,Font:11,Ocvirk:11,Lunnan:12,Rashkov:12,Simpson:13}.  
We take a somewhat different approach and
attempt to crudely peg the star formation rate in our model to the
measured rate in similar star forming environments in the local universe.

Specifically, evidence emerging from structural correlations is
suggesting that dwarf spheroidal galaxies are non-star-forming
analogs of dwarf irregular galaxies in the field \citep{Weisz:11,Kormendy:12,Kirby:13,Kirby:14}.  The dwarf spheroidals are
presently largely gas-free, likely because of stripping during the
infall into the host halo, that of the
Milky Way or Andromeda \citep[e.g.,][]{Grebel:03,Mayer:06}.  The
irregulars are still forming stars but the spheroidals ceased star
formation 
at some point in the past.  We exploit this evolutionary connection
and inform our modeling of star formation in the dwarf spheroidal progenitor objects
by the characteristics of star formation in low-surface-density and 
low-metallicity H\,I-dominated disks in the nearby universe.  This
allows us to question the necessity of explosive gas expulsion from
dwarf spheroidals, which is occasionally invoked to explain the properties of
at least some members of the dwarf spheroidal family \citep[e.g.,][]{Read:06,Sawala:10,Font:11,Kirby:11b}. 

This work is organized as follows.  In Section \ref{sec:histories}, we
use dynamical measurements of the dwarf spheroidals' central
mass densities to 
constrain their dark matter mass assembly histories.  With such histories
at hand, in Section \ref{sec:reionization} we utilize numerical calibrations of the impact of
reionization to estimate the evolution of the baryonic gas fraction in
the dwarf spheroidals' progenitor objects.  In Section
\ref{sec:star_formation_in_dwarfs}, we describe star formation in the
progenitor objects, arguing that they formed stars relatively
quiescently in what were
partially rotationally supported atomic gas flows.  In Section
\ref{sec:results}, we present our results which include estimates of
the stellar mass-radius and stellar mass-metallicity
relations.  Comparing these with the observed dwarfs allows us to estimate the masses and the formation
redshifts of the dark matter halos hosting the dwarfs.  In Section
\ref{sec:true_fossils}, we briefly discuss 
where the true pre-reionization fossil stellar systems might be
found in the Local Group, and in Section \ref{sec:conclusions}, we review
our main conclusions.  We delegate a number of minor clarifications
in instances where the prescriptions we employ diverge from those in
the literature to the footnotes.

\section{Dark Matter and Mass Assembly}
\label{sec:histories}

\subsection{The Common Mass Scale Halos Hosting Dwarf Spheroidal Satellites}
\label{sec:common_mass_scale}

Dynamical estimates of the masses enclosed within $300\,\textrm{pc}$
of the galaxy center in the Local Group dwarf spheroidals with luminosities
$\lesssim 10^7\,L_\odot$ seem to be consistent
among most of the dwarfs, suggesting a ``common
mass scale'' of 
\beq
M_{300}\approx 10^7\,M_\odot 
\eeq 
within this
radius \citep[][hereafter S08]{Strigari:08}. This mass seems independent of the 
galaxy luminosity, though there is a significant scatter, especially
at low luminosities, where the enclosed mass is measured at 
radii smaller than $300\,\textrm{pc}$ 
and the value of $M_{300}$ is then estimated by extrapolation assuming a density
profile of the NFW form \citep{Navarro:96}.    Such extrapolation
is meaningful in principle because dark matter strongly dominates the
gravitational potential in dwarf spheroidals, especially at low
luminosities.
In what follows, we refer to the
relation $M_{300}=10^7\,M_\odot$ as \emph{the} common mass scale family, but
we also carefully explore the dependence of our results on the variation of
$M_{300}$.  This additional degree of freedom allows us to classify all dark matter halos by their
value of $M_{300}$ and thus discuss alternative common mass scale
families that are more or less dense than the one identified in S08.

Measurements of the dwarf spheroidals' dynamical masses can be used to
place a simple constraint on the mass and epoch of their mass assembly
histories.
The density profile of an isolated
 halo evolves continuously as the halo accretes matter and merges with
 other halos.  In this fashion, halos grow in radius and 
 mass and their concentrations evolve.  The concentrations of low-mass
 halos corresponding to low-$\sigma$ cosmic peaks increase with
 time because these halos grow slowly, allowing for a low-density
 envelope to accrete around a denser core \citep{Wechsler:02}.   Since the
 radius of $300\,\textrm{pc}$ is much smaller than the virial radius
 of galaxy-hosting halos at the present epoch, the mass contributing to $M_{300}$
 must have collapsed at an early epoch, when the universe was denser.

The mass assembly of a dwarf spheroidal host halo is truncated by the non-linear
tidal field at a critical time preceding its eventual infall into
 a larger halo, that of the Milky Way, Andromeda, or one of their
main progenitors.  Recent numerical
 simulations of CDM clustering suggest that the mass
 assembly of a small dark matter halo is truncated when the halo
 approaches to within $\sim 3$ virial radii of the larger
 host halo (R.~Wechsler, priv. comm.).  At $\sim1.5$ virial radii of
 the larger halo, tidal stripping starts reducing the mass of the
 smaller halo \citep[e.g.,][see also
 \citealt{Wang:07,Dalal:08}]{Hahn:09}. 
The subhalo's
 outer layers are stripped but its densest central cusp can remain intact,
 a frozen-in relic of an earlier epoch, perhaps substantially preceding the
 infall.  

The development that follows applies to dwarf spheroidal satellites with
 such preserved central cusps. Our operating assumption is that the
 tidal stripping has not removed significant stellar mass from the
 dwarf spheroidal progenitor object.  Exceptions include the handful of examples, such
 as Sagittarius \citep{Belokurov:06a}, Hercules \citep{Coleman:07},
 and Ursa Major II \citep{Munoz:10}, with evidence for 
 tidal stripping.   In the
 tidally stripped satellites, the mass estimates obtained from stellar kinematics provide lower
 limits on what the central densities of these objects could have been
 in the past.

The host halos of dwarf spheroidal satellites 
 ostensibly belonging in the common mass scale family should be 
 such frozen-in relics.  Their outer dark matter envelopes have long been
 stripped and their current gravitationally bound 
masses are substantially reduced compared to the maximum at the point
of the tidal truncation of mass assembly. We refer to the maximum mass of the
 dwarf-hosting halo just before tidal stripping as the maximum tidally
 truncated
 mass $M_{\rm TT}$ and denote the corresponding
 redshift with
 $z_{\rm TT}\geq 0$.

Anticipating our focus on star formation in the dwarf spheroidal
progenitor objects below, we note that the cessation of dark matter
accretion may be followed with ram pressure stripping of the residual gas in
the tidally truncated subhalo \citep[e.g.,][]{Mayer:06}, effectively
shutting off further star
formation.  We consider $z_{\rm TT}$ an upper limit to the ram pressure stripping redshift $z_{\rm
  ram}$, the latter being the minimum redshift at which
the dwarf formed new stars. 

Theoretical expectations for the dark matter halo density profile as a
function of mass and redshift in the standard $\Lambda$CDM cosmology
have been calibrated with cosmological simulations. We assume that the halo density profile is NFW with concentration $c$
and scale radius $r_{\rm
  s}=r_{200}/c$, where $r_{200}$ is the radius at which the mean
density inside the halo equals 200 times the critical density of the
universe.\footnote{The halo density profile is better described
with the Einasto profile \citep[e.g.,][]{DiCintio:13,VeraCiro:13}, but
the development presented here does not critically depend on the assumed profile.}
We adopt the calibration of the halo mass and redshift dependence of
the median halo concentration derived by \citet{Prada:12} for 
halos in the Bolshoi and MultiDark simulations \citep[but see the
comments in][]{Ludlow:12}. This parametrization is
written in terms of the rms fluctuation $\sigma(M)$  of the density
field linearly extrapolated to $z=0$.  In evaluating $c(M,z)$ at low halo masses, we employ
$\sigma(M)$ computed from the best-fitting \citet{Planck:13} cosmological parameter
set [from the {\it Planck} temperature data and {\it Wilkinson Microwave
Anisotropy Probe} ({\it WMAP}) polarization] with the
help of the \textsc{camb} package \citep{Lewis:00}.\footnote{Equation 12  in \citet{Prada:12} giving the linear growth factor $D$ is not
  correctly normalized to unity at $z=0$.  We substitute the
  correctly normalized $D(a)$ into their Equation 23.  \citet{Prada:12} utilize the \citet{Klypin:11} fit
to $\sigma(M)$, which seems relatively accurate only for $M>10^8\,M_\odot$.}    We find that $\sigma(M)\approx \sum_{n=0}^3 C_k
\log^k(M/M_\odot)$ with $C_0=14.95$, $C_1=-0.9091$, $C_2=-0.04806$,
and $C_3=0.003031$ provides an excellent
fit in the mass range $10^5\,M_\odot\leq M\leq 10^{12}\,M_\odot$.

\begin{figure}
\begin{center}
\includegraphics[width=0.475\textwidth]{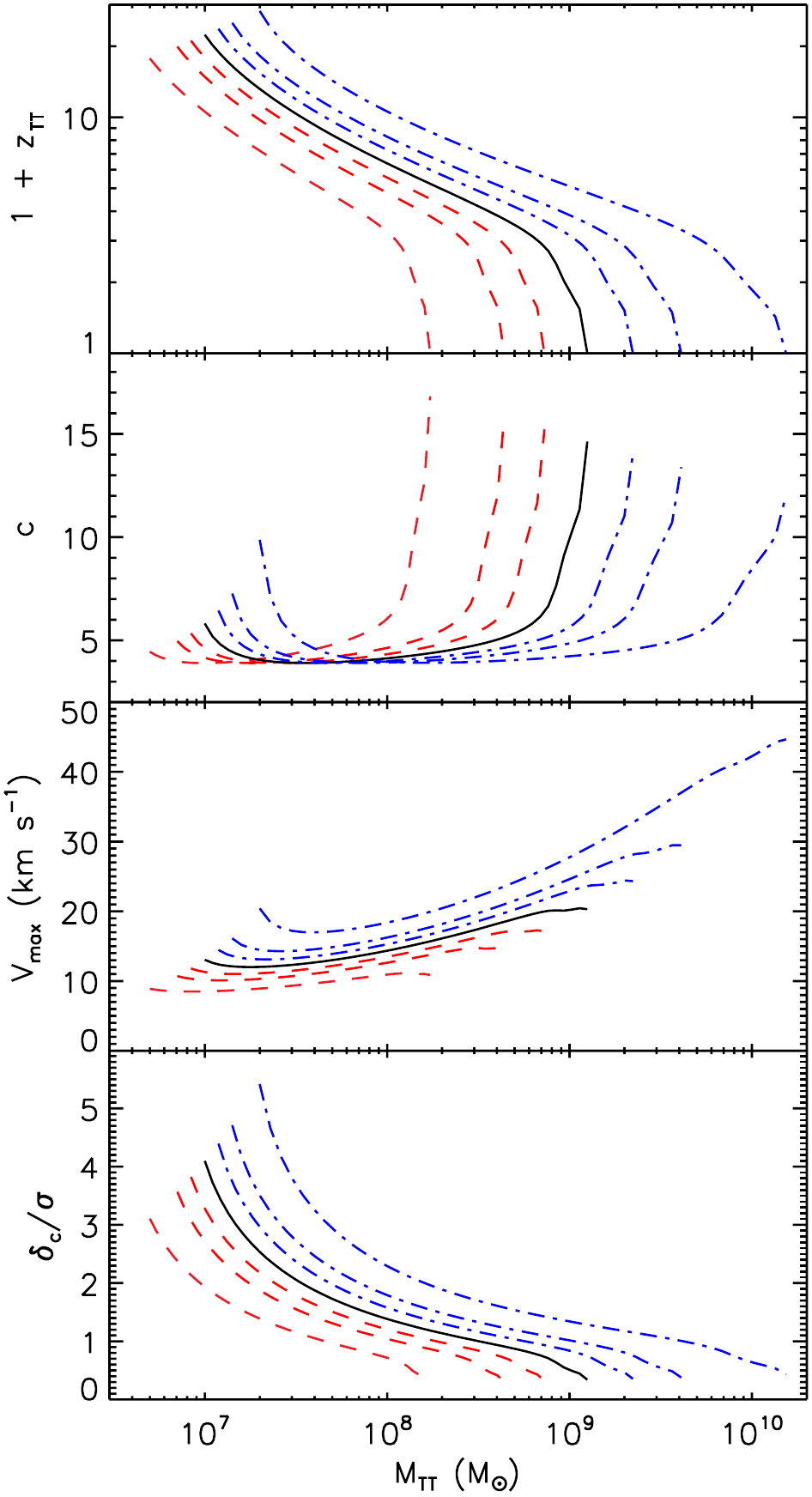}
\caption{Properties of the dark matter halo of mass $M_{\rm TT,0}$ belonging on
  the common mass scale relation $M_{300}=10^7\,M_\odot$ at the point at which the external tidal field
freezes the evolution of the inner density profile  (solid lines).  The panels from
top to bottom show the redshift
$z_{\rm TT,0}(M)$, the corresponding halo concentration $c$,
maximum circular velocity of the halo $V_{\rm max}$, and overdensity peak
height in units of the rms density variance $\delta_{\rm
  c}(z)/\sigma(M)$. The dashed and dot-dashed lines show
  the neighboring common mass scale relations as a function of
  $M_{{\rm TT},\epsilon}$ for
  $M_{300}=(2^{-1},\,2^{-0.5},\,2^{-0.25})\times10^7$ and
  $(2^{0.25},\,2^{0.5},\,2^{1})\times10^7\,M_\odot$, respectively.}
\label{fig:tidal}
\end{center}
\end{figure}

Halos belonging in a common mass scale family lie on the relation 
\beq
\label{eq:definition_CMS}
M_{\rm NFW} [300\,\textrm{pc}; M_{\rm TT},z_{\rm TT},c(M_{\rm TT},z_{\rm TT})]
= M_{300} ,
\eeq
where $M_{\rm NFW}(r; M_{\rm halo},z,c)$ is the mass enclosed within radius $r$
for an NFW halo of mass $M_{\rm halo}$ at redshift $z$ having concentration $c$.
For a specific value of $M_{300}$, 
this defines a one parameter family of halo masses evaluated at the point of
the tidal truncation of mass assembly
and the corresponding redshifts.  Depending on whether we treat the
redshift or the mass as an independent variable, we denote this family with
\beq
M_{{\rm TT},\epsilon} (z_{\rm TT}) , ~ ~ ~ ~ ~ ~ z_{{\rm TT},\epsilon}
(M_{\rm TT}) ,
\eeq
where the parameter $\epsilon$ in the subscript indicates the
departure of the 
$M_{300}$ parameter from the S08 value
\beq
\epsilon \equiv \log \left(\frac{M_{300}}{10^7\,M_\odot}\right) .
\eeq
Note that Equation (\ref{eq:definition_CMS}) also defines a maximum halo
 mass corresponding to $z_{\rm TT,\epsilon}=0$ that is consistent with the
 common mass scale
 relation.  In Figure \ref{fig:tidal}, we show $z_{\rm
   TT,\epsilon}(M_{\rm TT})$
 for several representative 
choices of $M_{300}$. For the S08 common mass scale family,
solutions with nonnegative tidal redshifts
can be found for $M_{\rm TT,0}\lesssim 1.25\times10^9\,M_\odot$, but this upper limit is
extremely sensitive to the central density and increases by an order
of magnitude to $M_{\rm TT,+0.3} \lesssim 1.5\times10^{10}\,M_\odot$ after doubling the central density.  We find that $z_{{\rm
  TT},\epsilon}$ increases steeply with decreasing $M_{\rm TT}$
until it reaches $z_{\rm
  TT,\epsilon}\sim 2$, and at lower masses and higher redshifts, the redshift increase is
more gradual.\footnote{The redshifts $z_{\rm TT,0}$ seem
systematically lower than those computed by \citet{Maccio:09} who find
$z_{\rm TT}(10^8\,M_\odot)\sim 7$ and $z_{\rm
  TT}(10^9\,M_\odot)\sim 2.5$. The discrepancy could arise
from differences in halo structural properties: derived from relatively
coarse resolution cosmological
$N$-body simulations in \citet{Maccio:09} and from analytical NFW profiles
based on the \citet{Prada:12} concentrations in the present work.} 

Figure \ref{fig:tidal}
further shows that the halo concentrations are approximately
independent of the halo mass, $c\approx 4$, over most of each common
mass scale family, but then increase steeply with increasing mass, to
$c\sim 10$ or higher, near the high-mass end of the common mass scale family,
where the corresponding redshifts drop below $z_{\rm
  TT,\epsilon}\lesssim 1$.  At these low redshifts, the dimensionless density
peak height given by the critical linear overdensity for collapse in
units of the $z=0$ rms density fluctuation, $\delta_{\rm c}(z_{{\rm
  TT},\epsilon})/\sigma(M_{{\rm TT},\epsilon})$, drops below unity, indicating
that the massive dwarf spheroidals formed in the collapse of very
low-$\sigma$ peaks in the cosmic density field.

The
maximum circular velocities of the common mass scale halos, also shown in
Figure \ref{fig:tidal}, increase very slowly with $M_{\rm TT,0}$ and cover the range
$10\lesssim V_{\rm
  max} \lesssim 20\,\textrm{km}\,\textrm{s}^{-1}$. 
Higher maximum circular velocities of $V_{\rm
  max}\gtrsim 30\,\textrm{km}\,\textrm{s}^{-1}$ require
$M_{300}\gtrsim 2\times 10^7\,M_\odot$.  The inability of the
S08 common mass scale family to accommodate high $V_{\rm
  max}$ satellite halos is a manifestation of the ``too-big-to-fail'' problem \citep{BoylanKolchin:11,BoylanKolchin:12}, which we briefly
discuss in Section \ref{sec:true_fossils}.

\subsection{The Universal Mass Accretion History of the Common Mass
  Scale Halos}
\label{sec:universal_history}

A crude sense about the formation histories of the common mass scale
halos can be obtained by estimating the average masses of the most massive
progenitors of halos belonging in the common mass scale one parameter
family.  Fortunately, accurate approximations describing the halo mass
growth rates are available. \citet{Fakhouri:10b} measured halo growth rates in the Millennium
CDM simulations and obtained a particularly accurate fitting function for
mean most massive progenitor
histories of halos in the mass range $10^{10}\leq M_{\rm halo}\leq
10^{14}\,M_\odot$.  The fitting function in their Equation 2 can be written in the form
\beq
\label{eq:Fakhouri}
\left(\frac{d\ln M_{\rm halo}}{dz}\right)_{\rm MMP} \approx - 0.62
\left(\frac{1+1.11\,z}{1+z}\right)
\left(\frac{M_{\rm halo}}{10^{12}\,M_\odot}\right)^{0.1} .
\eeq
The cosmological parameters in the Millennium simulations and the
\citet{Fakhouri:10b} analysis are slightly different from that in the
present work, but we nevertheless adopt the redshift dependence of their
fitting function, as well as the normalization at the reference mass
of $M_{\rm halo}=10^{12}\,M_\odot$.  Based on the extended Press-Schechter
excursion set theory \citep{Lacey:93}, the mass dependence of the
growth rate is set by the linear matter density fluctuation power spectrum
and can be approximated via \citep[see, e.g.,][]{Neistein:06}
\beq
\label{eq:MMP_growth_mass_scaling}
\left(\frac{d\ln M_{\rm halo}}{dz}\right)_{\rm MMP} \propto \left|
\frac{d\sigma^2(M_{\rm halo})}{d\ln M_{\rm halo}}\right|^{-1/2} .
\eeq
The mass dependence of Equation (\ref{eq:MMP_growth_mass_scaling}) evaluated
in the neighborhood of the reference mass $M_{\rm halo}=10^{12}\,M_\odot$ is $d\ln |d\sigma^2/d\ln
M|^{-1/2}/d\ln M\approx 0.094$, in good agreement with the mass dependence in
Equation (\ref{eq:Fakhouri}), but the slope implied by Equation
(\ref{eq:MMP_growth_mass_scaling}) flattens towards lower
masses to reach $d\ln |d\sigma^2/d\ln
M|^{-1/2}/d\ln M\approx 0.043$ at $M_{\rm halo}=10^6\,M_\odot$.  Therefore, we
adopt the mass dependence from Equation (\ref{eq:MMP_growth_mass_scaling}) to settle
on the following form
\bea
\label{eq:MMP_growth_mass}
\left(\frac{d\ln M_{\rm halo}}{dz}\right)_{\rm MMP} &=& - 0.62
\left(\frac{1+1.11\,z}{1+z}\right) \nonumber\\ & &\times \left[
\frac{d\sigma^2/d\ln M\, (M_{\rm halo})}{d\sigma^2/d\ln M \, (10^{12}\,M_\odot)} \right]^{-1/2} .
\eea

\begin{figure}
\begin{center}
\includegraphics[width=0.475\textwidth]{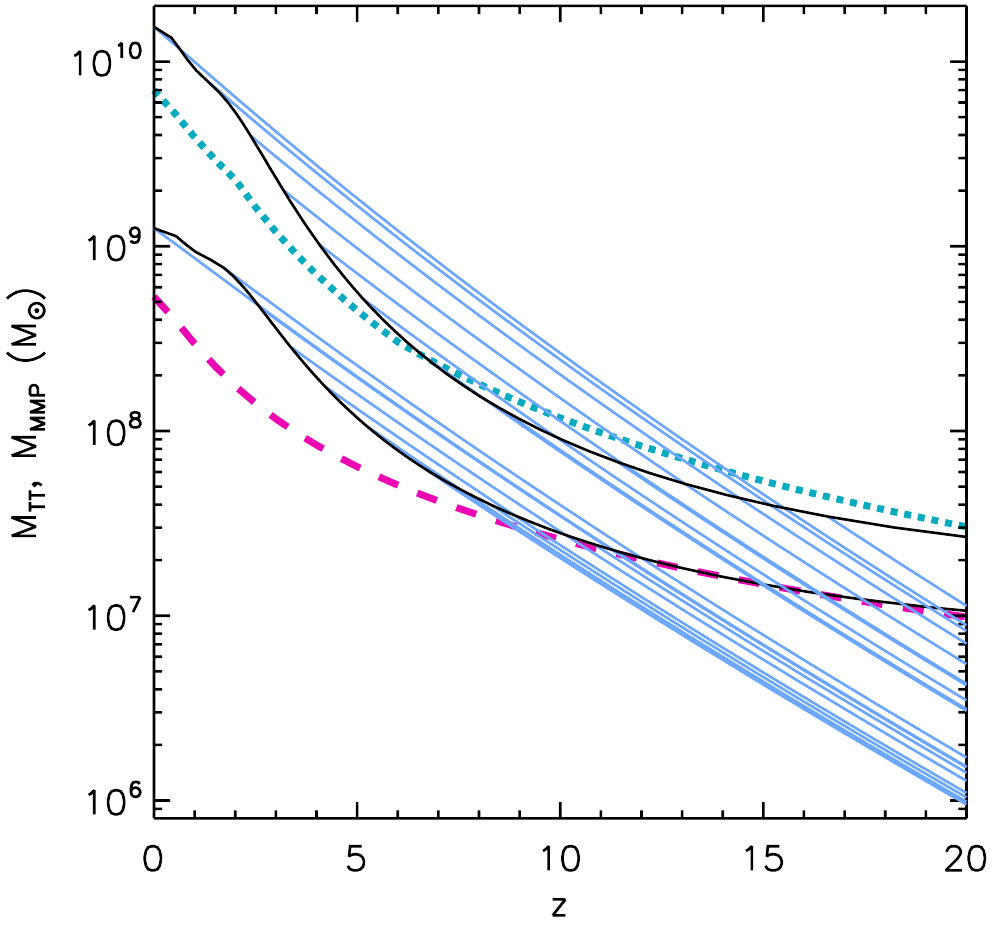}
\caption{Mass of the halos in the common mass scale family as a
  function of redshift $M_{{\rm TT},\epsilon}(z)$ for
  $M_{300}=10^7\,M_\odot$ (lower black curve) and
  $M_{300}=2\times10^7\,M_\odot$ (upper black curve).  The curves in
  light blue show representative most massive progenitor histories of the
  halos in the  two common mass scale families.  The thick dashed magenta
line and the thick dotted cyan line show the halo masses corresponding
to $T_{200}=10^4\,\textrm{K}$ (assuming neutral gas, $\mu_{\rm
  mol}=1.22$) and $T_{200}=T_{200,{\rm crit}}(z)$ (assuming ionized
gas, $\mu_{\rm mol}=0.6$) respectively (see Section \ref{sec:baryon_fraction}).}
\label{fig:mmp}
\end{center}
\end{figure}

Integrating Equation (\ref{eq:MMP_growth_mass}) we compute the mean most massive progenitor histories for halos
belonging in the common mass scale family which we denote with
\beq
M_{\rm MMP} [M_{\rm TT},z_{{\rm TT},0}(M_{\rm TT});z] .
\eeq
A sample of representative histories for two common mass scale
families, $M_{300}=10^7$ and $2\times10^7\,M_\odot$, is shown
in Figure \ref{fig:mmp}.  We find that the histories have a very weak dependence on $M_{\rm TT}$ at a
fixed redshift $z$, with the variation being particularly small for
the S08 family.  The common mass scale halos in the S08 family have
\emph{similar mean accretion histories}, at any given redshift varying
over less than a factor of 2 in mass. 
This can be understood as reflecting the fact that the dense material in the center
of a halo is put in place early on, only to evolve passively as the
mass and virial radius of the halo grow by accreting from a universe of
a progressively decreasing density.  The central densities of
the most massive progenitor halos evolve very little with redshift up to some maximum
redshift at which 
an active assembly of what is to become the dense central core of
the halo is taking place. This is consistent with the behavior
seen in Figure \ref{fig:tidal} in which the concentrations of halos with maximum tidally limited masses
$M_{\rm TT,0}\lesssim
10^8\,M_\odot$ are $c\approx 4$, approximately independent of the mass,
but at the highest halo masses, they increase sharply with the increasing
mass.

The universal history of the common mass scale halos is already
apparent in the conclusion of \citet{Wolf:10} that the dwarf
spheroidal satellites of the Milky Way, if they were to be artificially placed into
\emph{isolated} halos at $z=0$ as required to match their stellar
dynamical central density estimates, would all reside in similar mass
halos $M_{{\rm iso},z=0}\sim 3\times10^9\,M_\odot$.  This formal
conclusion is 
consistent within uncertainties with our estimate of the $z_{{\rm TT},0}=0$ common mass scale halo
mass. Nevertheless, given that the tidal truncation redshifts will vary
across the dwarf spheroidal satellite sequence, and so will the
average halo densities and concentrations, the physical significance of the $z=0$ isolated-halo-equivalent
masses of \citet{Wolf:10} is not entirely clear.

\subsection{Satellite Inspiral Times}

After a satellite halo becomes incorporated in the substructure of one
of the massive progenitors of the Milky Way or Andromeda, dynamical
friction drives its orbit to decay.  For the subhalo to be observed
as a satellite galaxy at $z=0$, the orbital decay time must be longer
than the time elapsed since the infall.  The orbital decay time depends
on the masses and concentrations of the
two halos and on the parameters of the orbit.  Here, we adopt a
crude, order-of-magnitude estimate of the orbital decay time
\citep[see, e.g.,][]{Cooray:05},
\beq
\tau_{\rm DF} \sim \xi \, \tau_{\rm ff} \, \frac{M_{\rm host}}{M_{\rm
    sat}} ,
\eeq
where $\xi$ is a dimensionless coefficient likely to lie in the range $0.1-1$, 
$\tau_{\rm ff}\sim [3\pi/(32\,G\bar\rho_{\rm host,sat})]^{1/2}\approx 0.1\,H(z)^{-1}$ is the
free fall time of the halos, $\bar\rho_{\rm
  host,sat}=200\times3H(z)^2/(8\pi G)$ is the average density of the halos,
$H(z)$ is the Hubble parameter at redshift $z$, and the host and
satellite halo masses can be set to 
\bea
M_{\rm host} &=& M_{\rm MMP} [M_{\rm MW},0;z_{\rm TT,0}(M_{\rm TT})] ,\nonumber\\
M_{\rm sat} &=& M_{\rm TT} .
\eea
The most massive progenitors of a Milky Way-like halo with a mass of
$10^{12}\,M_\odot$ at $z=0$ had average masses $\log (M_{\rm
  MMP,MW}/M_\odot)=(11.7,\,10.7,\,9.7)$ at redshifts $z=(1,5,10)$, respectively, in agreement
with Figure 6 of \citet{Fakhouri:10b}.  For the same redshifts, we
have $\log (M_{\rm TT,0}/M_\odot)\approx (9,\,8,\,7.3)$.  This shows
that $M_{\rm host}/M_{\rm sub}\sim 500$ for $M_{\rm TT,0}\gtrsim
10^8\,M_\odot$ and the ratio then gradually declines towards lower
masses and
higher infall redshifts.  Comparing the orbital
decay time to the Hubble time, we find
\beq
\tau_{\rm DF}\, H_0 \sim \xi\,\frac{50}{H(z)/H_0}\, \frac{M_{\rm host}/M_{\rm
    sat}}{500} .
\eeq
The value of $\xi$ must be constrained with high-resolution
cosmological simulations.  For example, for $\xi\sim 0.2$, the orbital decay time is longer than the Hubble
time, $\tau_{\rm DF}\,H_0> 1$, for halos of the S08 common mass
scale family with $z_{\rm TT,0}\lesssim 6$.  This
is sufficient to guarantee that a significant fraction of the S08
family progenitor objects will not have inspiralled too close to the 
Galactic Center and thus been completely disrupted.  For the
$M_{300}=2\times10^7\,M_\odot$ family, the maximum redshift for
avoiding orbital decay drops to $z_{\rm TT,+0.3}\lesssim 5$
(corresponding to $M_{\rm TT,+0.3}\gtrsim 5\times10^8\,M_\odot$),
implying that low-mass halos in the latter family will have been lost
to the Galactic Center.

\section{Reionization and Baryonic Content}
\label{sec:reionization}

\subsection{Dwarf Spheroidal Progenitors at Reionization}

It is widely considered that cosmic reionization defined the
properties of the dwarf spheroidal galaxy population in the Local
Group, but the
details of its role vary in the
rich literature on this subject \citep[e.g.,][]{Bullock:00,Benson:02,Grebel:04,Koposov:09,Munoz:09,Okamoto:09,Busha:10,Maccio:10,Font:11,Lunnan:12}.   The objects that formed
before reionization has swept through their local
protogalactic patches are sometimes called ``fossils'' \citep[e.g.,][]{Ricotti:05,Gnedin:06}.  The more
luminous, ``classical'' dwarfs in the Local Group contain stellar
populations spanning a wide range of stellar ages and are not fossils. The
ultra-faint dwarfs, however, are typically old \citep[e.g.,][and
references therein]{Brown:12}, and, from the standpoint of their stellar
ages and the statistics of their radial
distances from the center of their Local Group host galaxy,
are consistent with being fossils
\citep[e.g.,][]{Bovill:09,Bovill:11a,Bovill:11b,Munoz:09,Salvadori:09,Frebel:12}.  

It is striking that the ultra-faint and classical dwarfs seem to form
a single, one-dimensional sequence in the luminosity-radius-metallicity
space \citep[see,
e.g.,][]{Belokurov:07,Gilmore:07,Tolstoy:09,Wolf:10,Kirby:11a,Misgeld:11,Kormendy:12,McConnachie:12}, in addition
to having similar central mass densities. The latter property, of
course, places them on
the reported common mass scale relation.  This
continuity of properties raises the prospect of \emph{formation under
uniform conditions across the entire dwarf spheroidal satellite sequence}.
Reionization modifies these conditions drastically; therefore, it is
worth comparing the dwarfs' observed properties to theoretical
expectations for the properties of galaxies
forming under completely reionized conditions.  We proceed with this exercise,
leaving the question of the nature and manifestation in the Local Group
of the true pre-reionization fossils to a brief discussion in Section
\ref{sec:true_fossils}. 

What were the typical masses of the most massive
progenitors of the dwarf spheroidal satellite host halos at
redshifts at which the Local
Group could have plausibly undergone reionization?
For the common mass
scale halos in the S08 family, as Figure \ref{fig:mmp} shows, we find $M_{\rm MMP}\sim 10^7\,M_\odot$ at $z\sim 12-14$
and $M_{\rm MMP}\sim 10^8\,M_\odot$ at $z\sim 5-7$.  The reference
masses selected here bracket the range of the halo masses, known as the
``atomically cooling halos,'' in which the
baryons are first becoming able to cool by the collisionally excited Ly$\alpha$
line emission \citep[e.g.,][]{Oh:02,Bromm:11}.
The corresponding redshifts are low
compared to those of the first, Population III stars to form in the
Local Group, and bracket the range of redshifts at which reionization
of the Local Group volume is generally expected \citep[e.g.,][]{Li:13,Ocvirk:13}.  The
global reionization redshift of the universe as inferred from the analysis of
cosmic microwave background (CMB) anisotropy by artificially fixing the
reionization width to $\Delta z=0.5$ is $z_{\rm reion}=11.1\pm 1.1$
\citep[][the limits from {\it Planck} temperature data and {\it WMAP} polarization at
low multipoles]{Planck:13},
also belongs in this redshift range.  

The very first stars in the Milky Way already formed at
redshifts $\lesssim 35$ in halos
with masses $\gtrsim 10^5\,M_\odot$
\citep[e.g.,][]{Gao:10}.
As Figure \ref{fig:mmp} shows, however, the
mean most massive progenitors of the common mass scale objects
crossed the threshold for H$_2$ cooling, which enables 
metal-free star formation, at redshifts $\lesssim 20$ and at halo masses
$\gtrsim 10^6\,M_\odot$ \citep[e.g.,][]{OShea:07}.  At these
redshifts, the critical
halo mass for star formation will have possibly been substantially increased
by a growing H$_2$-molecule-dissociating
(Lyman-Werner) background \citep[e.g.,][]{Johnson:08,Ahn:09,Holzbauer:12,Fialkov:13}. At
redshifts $z_{\rm LW}\sim 15-20$, this background will
have raised the minimum mass of metal-free halos capable of forming
stars to a threshold $M_{\rm crit,LW}\gtrsim 10^7\,M_\odot$ at which the Ly$\alpha$
line cooling allows the gas to start collapsing to 
densities at which self-shielding from the dissociating radiation
becomes effective
\citep[e.g.,][]{OShea:08,Wolcott-Green:11,SafranekShrader:12}.  The
common mass scale objects' most massive progenitors will have crossed
$M_{\rm crit,LW}$ at redshifts at which the Local Group will have
already begun to become reionized.  

All this suggests that the dwarf spheroidal progenitor objects should have largely avoided forming
stars before reionization.
Some
may had been able to form a few stellar generations before becoming
affected by the UV backgrounds, e.g., in the
aftermath of a prompt enrichment by supernovae (SNe) from moderate-mass Pop
III stars \citep{Ritter:12}, but others would
have found themselves starless at the brink of reionization.
This is consistent with the continuity of structural and
chemical properties identified above.  We
note that
\citet{Koposov:09}, who modeled the Milky Way
satellite population with various prescriptions for baryonic mass
reduction after reionization and baryon-to-star conversion efficiencies, had previously
arrived at the same conclusion, that the best-fitting models require
that the bulk of the stars formed after reionization.

One useful simplification is to
encapsulate the impact of the ionizing background into a single
variable, the baryon fraction $f_{\rm b}$, which will normally be
limited from above by the cosmic mean $\Omega_{\rm b}/\Omega_{\rm
  m}$.  After a patch of the universe has been reionized and the baryons in small
halos in the patch photoevaporated \citep{Barkana:99,Shapiro:04}, the baryon
fraction, and with it the
threshold for star formation, is modulated by the 
thermodynamics of the photoionized intergalactic medium (IGM) and the relative strength
of the
gravitational and pressure forces
during halo assembly, as well as by the competition of ionization and
recombination in the densest gas located near the center of the halo
\citep[e.g.,][]{Thoul:96,Kepner:97,Kitayama:00a,Kitayama:00b,Kitayama:01,Dijkstra:04,Susa:04,Mesinger:08,Sobacchi:13a}.
The pressure force resists the dark matter's gravitational pull
already at the turnaround point of the gravitational collapse and thus
acts to
reduce the baryon mass fraction in virialized dark matter halos.  The pressure
is determined by the thermodynamic evolution of the gas, which is itself
a function of the detailed history of halo mass assembly, of
the chemistry of the gas, and of the
character of UV and X-ray radiation backgrounds.  
The interplay of these factors renders the problem of determining the
threshold for runaway baryonic collapse in reionized halos complex and
best addressed with cosmological
hydrodynamic simulations
\citep[e.g.,][]{Gnedin:00,Gnedin:12,Hoeft:06,Okamoto:08}.  
We proceed to model the
baryon fraction of the dwarf spheroidal progenitor halos aided with the results
of these numerical investigations.

\subsection{Baryon Fractions}
\label{sec:baryon_fraction}

The principal structural property of a halo determining
its baryon fraction is the depth of
the gravitational potential well, which can be 
quantified with a characteristic velocity or (virial) temperature. The velocity
or temperature can be compared
to the values required
 for
baryon retention in halos exposed to an ionizing background of a given
intensity.  
 A number of investigations have sought to calibrate
the dependence of the baryon fraction on halo
properties and the reionization history.  \citet[][hereafter O08]{Okamoto:08} measured
the baryon fraction in cosmological gas dynamical
simulations of halo collapse in a section of the universe undergoing reionization, and found that
the characteristic virial temperature $T_{\rm vir,crit}$ for a halo to
retain half of its universal allotment of baryons is approximately
independent of redshift
$T_{\rm vir,crit}\approx2\times10^4\,\textrm{K}$ at low redshifts $z\lesssim 2$ and
decreases steeply with increasing redshift at $z>2$.  The steep drop
is a consequence of the finite time, of the order of the sound crossing
time in the photoionized gas, that it takes gas to escape
the host
halos upon reionization, and also a consequence of the retention of dense
gas in infalling
subhalos with deep potential wells (T.~Okamoto, priv.~comm.). We expect that the critical virial
temperatures would have been higher than those estimated by O08
had unbound gas been excluded in the computation of baryon fractions. O08 neglected radiative transfer
effects such as self-shielding which could have had the opposite
effect.

 \citet[][hereafter SM13a]{Sobacchi:13a}
carried out spherically symmetric simulations of baryonic collapse in halos
before the completion of reionization, at $z\geq 6$, and computed the critical halo masses
$M_{\rm crit}$ for retaining half of the baryons, but now excluding
any unvirialized or unbound gas.  The resulting masses are
systematically higher than the corresponding masses in O08, perhaps a
consequence of the more selective criterion for tallying the
virialized baryons in SM13a,
but the redshift dependence is similar, again corresponding to a $T_{\rm
  vir,crit}$ that decreases with redshift.  SM13a fit
the critical mass in fully reionized halos to find a scaling
\beq
\label{eq:Sobacchi}
M_{\rm crit}\propto J_{\rm ion}^{0.17}\, (1+z)^{-2.1} ,
\eeq
where $J_{\rm ion}$ is the mean intensity of
the ionizing background. This fit was calibrated at $z\geq 6$ but we
extrapolate it to lower redshifts, where both UV and X-ray photons contribute to the
ionizations. The UV dominates until $z\sim 3$ and X-rays at
lower redshifts. Expressed in terms of the hydrogen photoionization rate
$\Gamma_{\rm HI}$, the background intensity has been measured to rise gradually
from $z\sim6$ to $z\sim5$ to a steady maximum level of $\Gamma_{\rm
  HI}\sim 10^{-12}\,\textrm{s}^{-1}$ \citep{Becker:13}. Then, after the peak of quasar
activity at $z\sim 2$, the photoionization rate declines sharply, by a
factor of $\gtrsim 10$ by $z=0$ \citep[][and references
therein]{Haardt:12}. 

It is not clear if the low-redshift drop in
$J_{\rm ion}$ should translate into a reinvigorated infall of baryons
into isolated
dwarf galaxy halos and whether that would lead to a corresponding decrease in $M_{\rm crit}$.  If some
reinvigorated infall does happen before the tidal
truncation of mass assembly, that
could give rise to renewed star formation in 
dwarf spheroidal progenitor objects \citep[e.g.,][]{Ikeuchi:89,Babul:92,Ricotti:09}.
However, this effect does not seem to be manifest in the results of
O08.
The reinvigorated infall may not be taking place because it takes a
finite time, similar to the Hubble time, for the baryons to cool and
fall back into the halos \citep{Noh:14},
especially given that additional hydrodynamic effects (shocks and bulk
flows; \citealt{Mo:05,BenitezLlambay:13}) associated with the
collapse of long-wavelength large scale structure 
modes are stirring and raising 
the entropy of the IGM at the relevant redshifts
and mass scales.  For this reason, we assume in agreement with O08 
that $T_{\rm vir,crit}$ is a constant or monotonically decreasing
function of $z$, not exhibiting a drop towards low redshifts that would
be naively expected from a falling $J_{\rm ion}$.

Before the completion of reionization, at $z>6$, the globally
averaged ionizing background intensity is expected to decline with redshift with the
decrease of the ionized volume fraction.  The local intensity inside cosmic
H\,II regions, on the other hand, will exhibit little evolution
with redshift \citep[e.g.,][where in H\,II regions, the
ionization rate is approximately constant, $\Gamma_{\rm
  HI}\sim (2-4)\times 10^{-13}\,\textrm{s}^{-1}$ over the redshift
interval $7<z<15$]{Sobacchi:13b}.  In view of our hypothesis that the
dwarf spheroidal progenitor objects in the Local Group formed in reionized patches,
we adopt the latter photoionization rate for $z>6$.

Given these considerations, we construct a model for $M_{\rm
  crit}(z)$ as follows.  First, we choose the $z=0$ values of $M_{\rm crit}$ and
\beq
T_{\rm 200,crit}=\frac{\frac{1}{2} \mu_{\rm mol} m_p G M_{\rm crit}}{k_{\rm B}
r_{200,{\rm crit}}} ,
\eeq 
where $\mu_{\rm mol}\approx 0.6$ is the mean molecular weight
in units of the particle mass in ionized gas, to
equal those reported by \citet{Gnedin:12} and consistent with
O08, 
\beq
M_{\rm crit}(0)=7\times10^9\,M_\odot, ~ ~ ~ ~ ~ T_{\rm
  200,crit}(0)=2.8\times10^4\,\textrm{K} .
\eeq  
Then, as in O08, we
assume that at $z< 2$, the $M_{\rm crit}$ halos have redshift-independent 
virial temperatures.  At redshifts $2<z<6$, we
utilize the scaling in Equation (\ref{eq:Sobacchi}) with an effective
ionizing background (incorporating both the UV and X-ray contributions)
that decreases exponentially in $z$ by a factor of $3$ over this redshift range.   At
redshifts $z>6$, we continue applying Equation (\ref{eq:Sobacchi}),
but we assume that the ionizing background parameter $J_{\rm crit}$ in
ionized patches is constant.  Since for $z>2$, $T_{\rm
  200}\propto M_{\rm halo}^{2/3} \,(1+z)$ is a good approximation, we have 
\bea
\frac{T_{\rm 200,crit}(z)}{T_{\rm 200,crit}(0)}=\begin{cases} 1  & , ~ ~
  z < 2 ,\\
  3^{-0.17 \,(z-2)/4} [(1+z)/3] ^{-0.4}  & , ~ ~ 2 < z < 6 , \\ 
3^{-0.17} [(1+z)/3]^{-0.4}  & , ~ ~ z > 6 .
\end{cases}
\eea

It remains to model
how, at a given redshift, the baryon fraction $f_{\rm b}(M_{\rm halo},z)$ varies with $M_{\rm
  halo}$ such that $f_{\rm b}(M_{\rm crit},z) =
\frac{1}{2}\Omega_{\rm b}/\Omega_{\rm m}$.  Several models describing the
variation of the baryon fraction with the ratio of the halo mass to
the critical mass $M_{\rm halo}/M_{\rm crit}(z)$, but otherwise having
no explicit dependence on redshift, exist in the literature. We 
utilize the \citet{Gnedin:00} form 
\beq
\label{eq:f_b_Gnedin}
f_{\rm b}(M_{\rm halo},z) \approx \frac{\Omega_{\rm
  b}}{\Omega_{\rm m}} \left\{1+(2^{\alpha/3}-1)\, \left[\frac{M_{\rm halo}}{M_{\rm crit}(z)}\right]^{-\alpha}\right\}^{-3/\alpha} ,
\eeq 
where $\alpha$ is a shape parameter.\footnote{SM13a adopt a
  simpler functional dependence:
\beq
f_{\rm b}(M_{\rm halo},z) \approx \frac{\Omega_{\rm
  b}}{\Omega_{\rm m}} \frac{e^{-M_{\rm halo}/M_{\rm crit}(z) + 1}}{2} .
\eeq}
O08 report a good fit with 
$\alpha=2$, whereas
\citet{Gnedin:12}, who has also measured the baryon fraction in
cosmological simulations, reports $\alpha=1$.  The two choices of
$\alpha$ imply substantial differences for
$M_{\rm halo}\sim M_{\rm crit}$ but both scale as $f_{\rm b}\propto
M_{\rm halo}^{-3}$ for
$M_{\rm halo}\ll M_{\rm crit}$.  We adopt the \citet{Gnedin:12} functional form with $\alpha=1$
since SM13a agree that it fits their data.

\begin{figure}
\begin{center}
\includegraphics[width=0.475\textwidth]{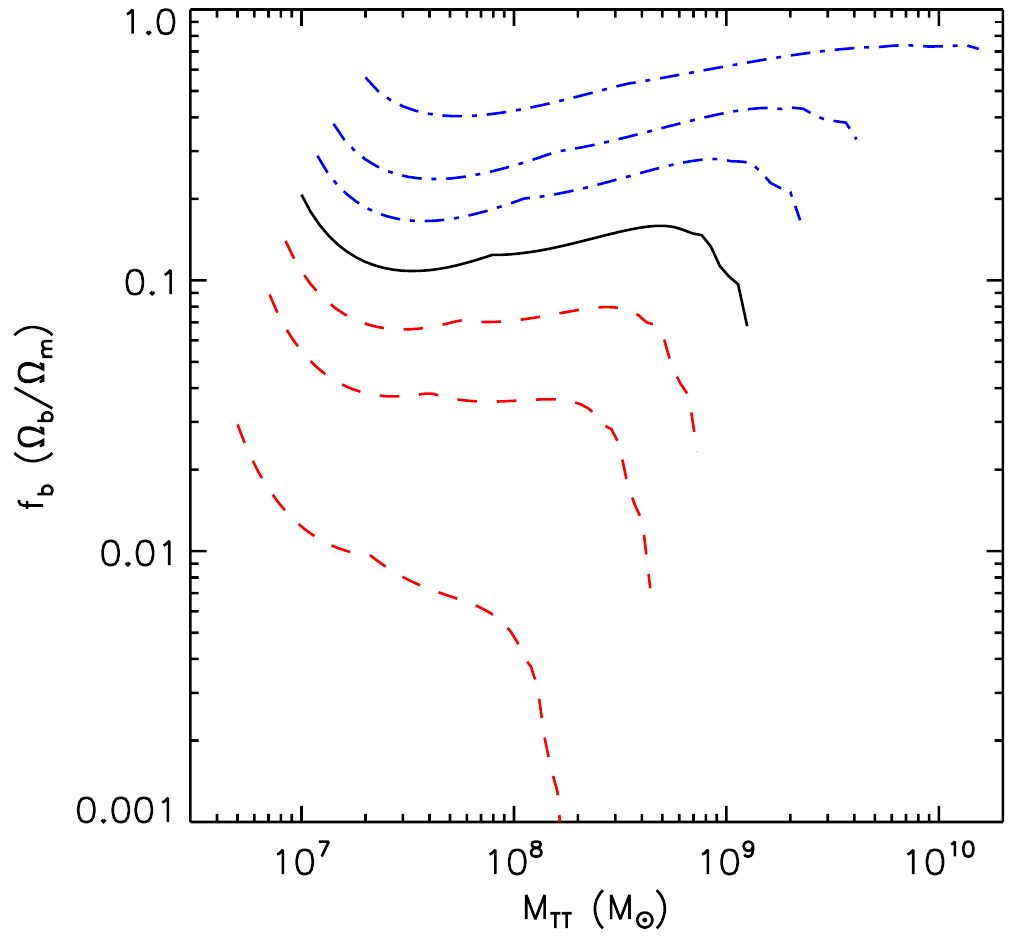}
\caption{The baryon fraction in units of
the universal baryon fraction $(\Omega_{\rm b}/\Omega_{\rm m})^{-1}$
at the point at which the external tidal field
freezes the evolution of the inner density profile  (black solid line).
The red, dashed and blue, dot-dashed lines show
  the neighboring common mass scale relations for
  $M_{300}=(2^{-1},\,2^{-0.5},\,2^{-0.25})$ and
  $(2^{0.25},\,2^{0.5},\,2^{1})\times10^7\,M_\odot$, respectively.}
\label{fig:baryon}
\end{center}
\end{figure}

In Figure
\ref{fig:baryon}, we show the baryon fractions of the common mass scale
halos at the point of the tidal truncation of mass assembly.   They are approximately independent of
the halo mass and are all substantially below the cosmic mean, $f_{\rm
  b}\sim (0.1-0.15)\,\Omega_{\rm b}/\Omega_{\rm m}$. The
small variation of
$f_{\rm b}$ with $M_{\rm TT,\epsilon}$ seen in Figure
\ref{fig:baryon} is not significant in
view of the various model uncertainties.\footnote{The baryon fractions are
somewhat larger than the values $f_{\rm b}\sim (0.01-0.05)\,\Omega_{\rm
  b}/\Omega_{\rm m}$ found by \citet{Simpson:13} in cosmological
simulations modeling the formation of dwarf spheroidal satellite-like
galaxies.}  The scaling of the baryon fraction with a single parameter,
$M_{300}$, if not a modeling coincidence, seems interesting in itself,
perhaps suggesting that it is the free fall time of the inner halo
(which is directly related to the density and thus to $M_{300}$) that
determines the degree of reionization-induced baryonic infall suppression.

Because of the steep
decrease of the baryon fraction with decreasing halo mass at a fixed
redshift, it is clear that any halos that are less dense in their
central $300\,\textrm{pc}$ than the common mass scale halos would have
great difficulty retaining even a small fraction of their universal allotment of
baryons and, consequently, would not be expected to form any stars.
One may thus be tempted to conclude that the common mass scale is shaped by
the influence of reionization: halos that are denser than those on the
common mass scale entered a rapid star formation mode before
reionization and leave behind stellar systems that would not be
recognized as dwarf spheroidals---the ultra-faint and classical dwarfs---in the Local Group.  We speculatively revisit the
question of the nature of pre-reionization fossils in Section \ref{sec:true_fossils} below.
On the other hand, objects that are less dense than those on the
common mass scale retained too small a fraction of their baryons to
form stars, and remained dark (see also Section \ref{sec:edges} below,
where we find that gas in the dwarfs with $M_{300}\lesssim
5\times10^6\,M_\odot$ may remain in the ionized phase).

Other processes may additionally influence the
baryon fraction in low-mass halos, such as gas heating \citep[][however, see
\citealt{Crain:07}]{Mo:05} and ram pressure stripping
\citep{BenitezLlambay:13} by shock waves and flows produced by the
collapsing large-scale structure, as well
as the expulsion of baryons by the feedback from star formation in the
host halo \citep[e.g.,][]{Okamoto:10,Font:11}.  The efficiency and
statistics of these processes are still highly
uncertain. Additional sources of theoretical uncertainty that are
particularly challenging to remove, even with the best currently available 
cosmological hydrodynamical
simulations, include the transport of nucleosynthetic products from
their progenitor stars (including the very first, Population III)
which directly affects the thermodynamic evolution of star-forming gas, and the
potential survival of compact, dense, pressure-confined clouds in the
face of the
disruption by reionization and star formation feedback.
Our intention here is to develop a minimal model for the
formation of the dwarf spheroidal satellite population in the Local Group, and
thus we do not attempt to characterize the impact of these processes.

\section{Star Formation}
\label{sec:star_formation_in_dwarfs}

\subsection{How Did Dwarf Spheroidal Satellites Form Their Stars?}

We turn to modeling star formation in the dwarf spheroidal progenitors.
First clues about the character of the gas flows that formed the bulk
of the stars can be obtained from the galaxy structural properties.
The dwarf spheroidal satellite stellar systems form a one-dimensional family in
the luminosity-effective radius or stellar mass-effective radius plane
\citep[][]{Belokurov:07,Gilmore:07,Tolstoy:09,Wolf:10,Kirby:11a,Misgeld:11,Kormendy:12,McConnachie:12}. The family is well separated
from the families occupied by the more concentrated
globular clusters, ultra-compact dwarfs, and dwarf elliptical galaxies
(M32), but connects
smoothly with the family of spheroidal galaxies \citep{Kormendy:09}.
Excluding the stellar nuclei that are ubiquitous in the comparatively
luminous 
spheroidals \citep[e.g.,][]{Cote:06}, and are seen in  
Sagittarius and NGC 205 in the Local Group, surface brightness profiles of the entire spheroidal family have
low S\'ersic-indices $n\sim 1-2$, similar to the index of the exponential
disk $n=1$.  

The relation defined by
the dwarf spheroidals' stellar masses and effective radii also
agrees with
that of the
galaxy disks \citep[e.g., Figure 13 of][]{Gadotti:09} and of the disky, low-luminosity dwarf irregular galaxies in the range of stellar
masses where the two morphological types overlap \citep[see,
e.g.,][]{Kormendy:12}. The luminosity- (or stellar mass-) metallicity relation for dwarf
irregulars also coincides with that for dwarf spheroidals \citep{Kirby:13}.  The average star formation histories of the two
populations and of the intermediate ``transition dwarf'' population also
form a continuum \citep{Weisz:11}. The disks themselves are
well described with the exponential surface brightness profile.  The structural similarity
between the two populations seems to suggest a disk-like origin for
the dwarf spheroidals, in the sense that the gas in which the
stars formed was, as in disks, rotationally supported
and endowed with an angular momentum distribution that would produce an
exponential surface density profile.  The observation that the
lowest-luminosity galaxies in group and cluster environments and in the
field are
almost exclusively spheroidal and dwarf irregular, respectively (A.~Klypin, priv.~comm.), further supports the galaxy transformation
paradigm \citep[e.g.,][]{Grebel:03} in which the lowest mass galaxies
transition through the star-forming,
dwarf irregular form before getting accreted on to a more massive
system and stripped of the gas.

The disk-like gas flows, which we will
argue have produced the dwarf spheroidal satellites in the Local
Group, can be
contrasted with the unstable, disordered flows seen in galaxy
mergers and in systems containing high-accretion-rate,
high-Mach-number baryonic inflows (sometimes called ``cold-mode'' accretion).  The disordered
flows are expected and observed to exhibit rapid gas transport into the central region or pervasive, large-scale clumping.  It is believed that disordered
flows produce more concentrated central stellar components similar to the stellar
bulges and nuclear clusters in disk galaxies and stellar density
cusps in ellipticals \citep{Hopkins:09}.  Disordered flows arise when gas is globally self-gravitating, 
or is flowing into the galaxy with highly supersonic
velocities, or has somehow lost much of its
angular momentum, e.g., in the fluctuating gravitational field of a
galaxy merger.

The dwarf spheroidal progenitor objects, if forming in a
reionized patch of the universe, do not seem to
fulfill the conditions for the development of violently disordered flows. Their reduced baryon
fractions imply that the gas condensing in these halos is unlikely to
have been globally
self-gravitating; the baryons settle in rotating hydrostatic 
quasi-equilibria dominated by dark matter gravity.  Their analogs
among the field galaxies in the
nearby universe are the H\,I-rich (typically ``irregular'') dwarf galaxies
\citep[e.g.,][]{Begum:06,Roychowdhury:10,Cannon:11,Papastergis:11,Huang:12},
which exhibit varying degrees of rotational and pressure support and
very little or no evidence for molecular gas. The paucity of molecular
gas is 
consistent with the dwarf galaxies having relatively low total gas
surface densities, below the limit for efficient atomic-to-molecular conversion.

Our picture is that the dwarf spheroidal progenitor
objects formed the bulk of their stars over extended periods in partially rotationally
supported, globally gravitationally stable
neutral gas flows, which we will refer to as ``disks.''  In halos hosting such disks, continued gas accretion from the
cosmic web could induce directional drift of the angular momentum axis
of the disk. Stars would inherit the instantaneous sense of rotation
of the gas flow
at the time of formation, and this sense could evolve over time.  
Collisionless dynamical processes, such as the tidal stirring during the infall of the
dwarf galaxy into a more massive halo \citep[e.g.,][]{Mayer:01a,Mayer:01b,Kazantzidis:11},
might further modify the stellar orbital structure and erase coherent rotation. The resulting stellar systems forming
from disk-like flows would no longer have
disk-like kinematics; they might in fact end up ``dynamically hot,''
consistent with the spheroidal classification.  

We
will assume that similar to the 
low-redshift H\,I disks in the spirals \citep[e.g.,][]{Swaters:02}
and dwarf irregulars \citep[e.g.,][]{Hunter:11}, the dwarf spheroidal
progenitor disks had exponential gas surface density radial profiles.  
A corpus of observational and theoretical investigations of star formation has
isolated the gas column density and metallicity as the primary
parameters controlling the star formation rate.  At column densities
above a threshold of 
$\gtrsim 10\,M_\odot\,\textrm{pc}^{-2}$ at metallicities $Z\sim Z_\odot$ and
potentially higher thresholds
at $Z\ll Z_\odot$, the disk gas is predominantly in
the molecular phase; at lower column densities, it is primarily atomic.
We will find that reduced baryon fractions (Section
\ref{sec:baryon_fraction}) will imply relatively low total gas column
densities in the dwarf spheroidal progenitor disks, suggesting
predominantly atomic gas.  

Many of the details of star formation in the regime in which
H\,I dominates the chemical state of the gas
remain to be understood, but some trends are emerging from recent
work.  The heating and cooling
processes in the gas should be strong functions of metallicity, but
the net effect of the metallicity on gas thermodynamics may involve competing
influences that partially cancel each other \citep[see, e.g.,][]{Krumholz:11,Glover:12a,Glover:12b}.
The
heating at typical densities characteristic of the warm neutral
medium (WNM) to cold neutral medium (CNM) transition, which is 
itself driven by the thermal instability \citep[e.g.,][and references therein]{Saury:13}, is
facilitated by the photoelectric effect on dust
grains.  The dust abundance increases with metallicity, as does the
ability of the gas to shield itself from the interstellar radiation field. At low
metallicities and reduced molecule abundances, the 
cooling by atomic fine structure lines at low densities and dust at
high densities dominates over molecular cooling.
While the star formation rate clearly correlates with the H$_2$
abundance inferred from CO observations \citep[e.g.,][]{Schruba:11}, the
molecules in the atomically dominated regime can be thought 
of as a tracer of the star-forming CNM \citep[e.g.,][]{Krumholz:11,Krumholz:12b}.

The average stellar metallicities of the dwarf spheroidals are low,
e.g., $-2.6\lesssim \langle[\textrm{Fe}/\textrm{H}]\rangle\lesssim -1$
\citep{Kirby:11a}, and these are lower metallicities than the ones
for which star formation rates as a function of H\,I and H$_2$
column densities have been calibrated with H$\alpha$ and
far-ultraviolet (FUV)
observations in the nearby universe.  Theoretical star
formation prescriptions that utilize the linear relation between the H$_2$
and star formation rate surface densities \citep[e.g.,][]{Krumholz:12a,Kuhlen:12,Kuhlen:13} are likewise not valid at
metallicities as low as found in the dwarf spheroidals, where
they would predict complete absence of star formation \citep[see,
e.g., the discussion in][where an artificial floor on the H$_2$
abundance is imposed]{Kuhlen:13}.  We do
not attempt to construct a theoretical model for the star formation
rates, but turn to the local dwarf irregular galaxies for clues about
the efficiency with which the dwarf spheroidal progenitor objects
formed their stars.

A number of measurements of star formation rates at low column
densities and low metallicities based on H$\alpha$ and FUV fluxes in the nearby dwarf
irregulars \citep[e.g.,][]{Begum:08,Roychowdhury:10,Roychowdhury:11}
are consistent with the time-scales for H\,I depletion by
star formation $\tau_{\rm dep,Irr} = \Sigma_{\rm
    HI}/\Sigma_{\rm SFR} \sim 10\,\textrm{Gyr}$.
Other measurements agree with or potentially exceed this scale.
\citet{vanZee:01} measured an average depletion time-scale of
$\tau_{\rm dep,Irr}\sim 20\,\textrm{Gyr}$ in a sample of dwarf
irregular galaxies.  The conditions in dwarf irregular gas disks
(surface densities, metallicities, tidal field strengths) are
approximately mimicked
in the outer disks of more massive disk galaxies.
\citet{Bigiel:10} carried out FUV measurements of the star formation rates in the outer disks of
the nearby disk galaxies and found gas depletion time-scales ranging
from $\tau_{\rm dep,edge}\sim 100\,\textrm{Gyr}$ at $\Sigma_{\rm HI}\sim
1\,M_\odot\,\textrm{pc}^{-1}$ to $\tau_{\rm dep,edge}\sim (10-30)\,\textrm{Gyr}$
at  $\Sigma_{\rm HI}\sim
10\,M_\odot\,\textrm{pc}^{-1}$.  \citet{Hunter:11}  find a similar
trend examining the outer edges of five dwarf irregulars.  At higher surface densities, the
depletion times drop below $10\,\textrm{Gyr}$ \citep{Schruba:11}.  
In view of these empirical constraints, we adopt the time-scale for H\,I-to-stars conversion of 
\beq
\label{eq:depletion_time}
\tau_{\rm dep} = \frac{M_{\rm HI}}{\dot{M}_\star} \sim 
10\,\textrm{Gyr} 
\eeq
as the star formation time-scale in the dwarf spheroidal progenitor
objects.\footnote{This depletion time is a factor of $\sim 5-10$
  shorter than the depletion times estimated by \citet{Kuhlen:13} for
  halos of mass $M_{\rm halo}\lesssim
10^{10}\,M_\odot$ at $z=2.5$. The success of our choice of $\tau_{\rm
  dep}$ in explaining the dwarf spheroidal population of the Local
Group (see Section
\ref{sec:results} below) suggests that reconciling the longer $\tau_{\rm dep}$ of
\citet{Kuhlen:13} with the observed luminosity-radius relation would
require an evolutionary reduction of the dark matter density as
quantified by $M_{300}$, e.g., by baryonic processes \citep[see, e.g.,][]{Arraki:12,Governato:12,Pontzen:12,Brooks:13,GarrisonKimmel:13}. }

\subsection{Gas Disks in Dwarf Spheroidal Progenitors}

\subsubsection{Surface Densities}

\begin{figure}
\begin{center}
\includegraphics[width=0.475\textwidth]{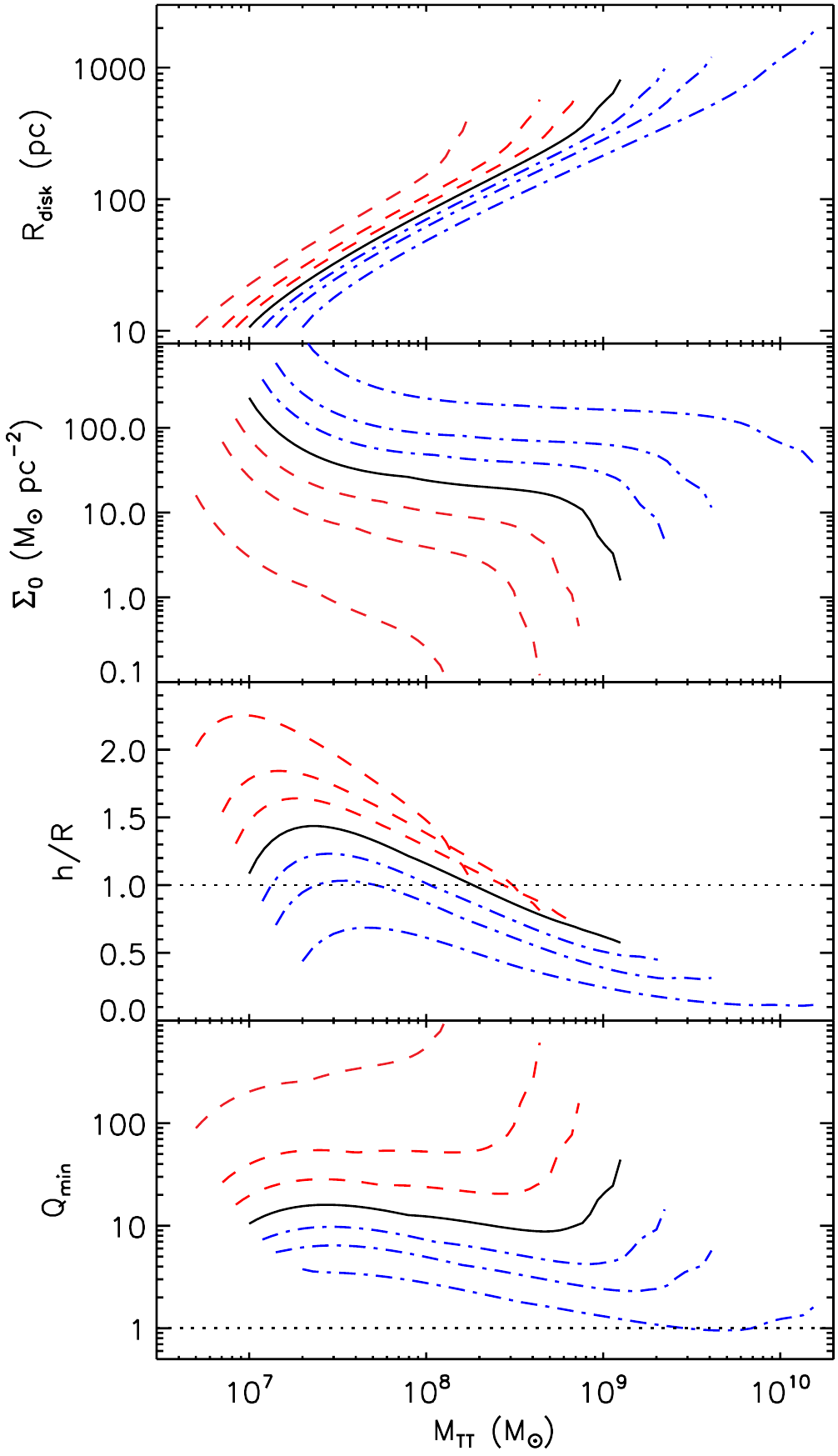}
\caption{Properties of the baryonic content of dark matter halo of
  mass $M_{\rm TT,0}$ belonging on
  the common mass scale relation $M_{300}=10^7\,M_\odot$ at the point at which the external tidal field
freezes the evolution of the inner density profile  (solid lines).  The panels from
top to bottom show the radius of the gas disk $R_{\rm disk}$, the central density of the gas disk $\Sigma_0$, the relative vertical disk scaleheight $h/R$ evaluated at
  $r=R_{\rm disk}$, and the
minimum instability parameter of the disk $Q_{\rm min}$.  In computing
these quantities, we assumed a spin
  parameter of $\lambda=0.05$ and a gas mass fraction in the disk of
  $f_{\rm disk}=0.5$.  The dashed and dot-dashed lines show
  the neighboring common mass scale relations for
  $M_{300}=(2^{-1},\,2^{-0.5},\,2^{-0.25})$ and
  $(2^{0.25},\,2^{0.5},\,2^{1})\times10^7\,M_\odot$, respectively.}
\label{fig:disk}
\end{center}
\end{figure}

Our approach to modeling the gas disks in the dwarf spheroidal progenitor
objects is based on the standard methods \citep[e.g.,][]{Mo:98,Schaye:04,Dekel:13}.
We assume that the radial dependence of the baryon surface density in the collapsed
component is 
exponential 
\beq
\Sigma(R) =\Sigma_0 e^{-R/R_{\rm disk}} ,
\eeq
consistent with the ubiquity of the exponential profile in outer H\,I
disks in nearby galaxies \citep[e.g.,][]{Swaters:02,Hunter:11}. Here,
  $R_{\rm disk}$ denotes the characteristic disk exponential scalelength, which can
  be written in terms of the dimensionless baryonic spin parameter $\lambda$
  via 
\beq
R_{\rm disk}= \frac{\lambda\, r_{200}}{\sqrt{2}} .
\eeq 
We take this
  relation, rather than the usual one expressing the angular momentum
  of the disk in terms of that of the halo, to define $\lambda$.  While $\lambda$
  is set by the linear and non-linear torques during gravitational
  clustering and hydrodynamical evolution and varies stochastically
  from galaxy to galaxy, here we treat it as a free parameter with typical values
  $\lambda\sim 0.05$ \citep[e.g.,][]{Dutton:11,Kravtsov:13}.\footnote{The value
    that \citet{Dutton:11} quote, $\lambda=0.035$, is defined
    relative to the virial radius $r_{\rm vir}$ which is somewhat
    larger than $r_{200}$.  \citet{Kravtsov:13} derives $R_{\rm
      disk}\approx 0.03\,r_{\rm 200}$ corresponding to $\lambda\approx
    0.04$.}  Disk radii of the dwarf spheroidal progenitor objects at the point of the tidal truncation of mass
  assembly 
  are shown in Figure \ref{fig:disk}.  They increase
  with the halo mass much more steeply, 
$R_{\rm disk}\propto (M_{{\rm TT},\epsilon})^{2/3}$ to $R_{\rm
  disk}\propto M_{{\rm TT},\epsilon}$, than a family of halos
  collapsing at the same redshift, which would have $R_{\rm
    disk}\propto M_{\rm halo}^{1/3}$. 

We allow that a
  fraction $f_{\rm disk}\sim 0.5$ of the baryonic content of the halo resides in
  the disk; thus $\int_0^\infty \Sigma(R) 2\pi R dR = f_{\rm disk}
  f_{\rm b} M$.  From this we find that the central surface mass density of the
  baryonic disk in the dwarf spheroidal progenitor objects is 
\beq
\Sigma_0= f_{\rm disk} f_{\rm b} \,\frac{M_{\rm halo}}{ 2\pi R_{\rm disk}^2} .
\eeq
 The resulting disk central surface densities are shown in Figure
 \ref{fig:disk}.  In the S08 common mass scale family with
 $M_{300}=10^7\,M_\odot$, 
surface densities are in the range $\Sigma_0\sim (10-100)\,M_\odot\,\textrm{pc}^{-2}$,
  sufficient for low-level star formation to proceed and yet insufficient, given
  the 
  low metallicities, for a substantial gas
  fraction to transition into molecular form. The central surface density
  is particularly sensitive to the halo central density parametrized
  with $M_{300}$, a
  consequence of the sensitivity of the baryon fraction to the
  halo central density (see Section \ref{sec:baryon_fraction}).  In the common mass scale families
  with $M_{300}\lesssim 0.5\times10^7\,M_\odot$, the surface
  densities are so low that these objects
 are utterly unable to form stars.   On the other hand, the objects with
 $M_{300}\gtrsim 2\times10^7\,M_\odot$ seem to have surface densities
 sufficient to form dense molecular clouds even at relatively low metallicities.

\subsubsection{Vertical Structure}
\label{sec:vertical_structure}

The relative degree of rotational and pressure support can be
estimated by evaluating the dimensionless 
ratio $h/R$ of the vertical pressure scaleheight of the disk to the
radius.  When $h/R\ll 1$, the disk is geometrically thin and radial
force balance is provided by rotation. When, at least formally,
$h/R\gtrsim 1$, the equilibrium is not a true disk but a geometrically thick atmosphere in
which radial force
balance is provided by the hydrostatic pressure.   The pressure scale
can be estimated as the ratio of the effective (thermal and turbulent
added in quadrature)
velocity dispersion of the disk to the natural frequency of
vertical oscillations, 
\beq
h\sim \frac{c_{\rm eff}}{\mu} .
\eeq
The square of the
frequency is given by the second derivative of gravitational potential
along the vertical direction, 
\beq
\mu^2=\left.\frac{\partial^2\Phi}{\partial
  z^2}\right|_{z=0} . 
\eeq
Both the halo and the disk contribute to the
  gravitational potential. The halo contribution and the corresponding
  vertical frequency $\mu_{\rm halo}$ can be
  straightforwardly computed
  from the NFW profile.   The gravitational potential near
  the surface of a 
  razor-thin disk is $\Phi_{\rm disk} =2\pi G\Sigma |z|$. Allowing
  for non-zero half-thickness $h$, we have $\mu_{\rm disk}^2=
  (\partial^2\Phi_{\rm disk}/\partial z^2)_{|z|\lesssim h} \sim 2\pi
  G\Sigma/h$.  Here, $h$ implicitly depends on both the halo and the
  disk component.  Solving for, $\mu=(\mu_{\rm halo}^2+\mu_{\rm
    disk}^2)^{1/2}$ we obtain
\beq
\mu = \frac{\pi G\Sigma}{c_{\rm eff}} + \sqrt{\left(\frac{\pi G\Sigma}{c_{\rm eff}} \right)^2+\mu_{\rm
  halo}^2} .
\eeq

For simplicity we assume that $c_{\rm eff}$ is independent of $z$; this is most certainly
not the case as the low-dispersion CNM will reside
closer to the midplane.  In dwarf
  irregular galaxies, $c_{\rm eff}$ varies from
  $\sim15\,\textrm{km}\,\textrm{s}^{-1}$ in the warm phase H\,I to
  $\sim 5\, \textrm{km}\,\textrm{s}^{-1}$ in the cold phase
  \citep[e.g.,][see also \citealt{Stilp:13} as well as the theoretical
  results in \citealt{Saury:13}]{Ianjamasimanana:12,Zhang:12}. The dwarf
  spheroidal progenitor disks with suppressed baryon fractions seem to have
  $\mu_{\rm halo}\gtrsim \mu_{\rm disk}$ for $\lambda\sim 0.05$, indicating that dark matter
  dominates vertical gravity.  

In Figure \ref{fig:disk}, we plot
  $h/R$ evaluated at $R=R_{\rm disk}$ for halos in the S08 common mass scale family and several
  neighboring families.  We find that the disks with $M_{{\rm
   TT},0}\lesssim 10^8\,M_\odot$ are typically thick with $h/R\gtrsim
  1$ for $R\lesssim R_{\rm disk}$. Since, approximately, $h/R\propto R^{-1/2}$, the
  disks become thin at larger radii.  Thus the inner radii of the gas
  disks can be pressure
  supported, especially towards lower halo masses. This resonates with
  the observation that the dwarf irregular galaxies seem to have solid-body like, $V_{\rm HI}(R)\propto R$
  rotation curves in the centers.  Such rotation curves are often
  interpreted as evidence for constant-density cores in the dark matter
  distribution, but could also arise simply due to pressure support at
  the innermost radii.

\subsubsection{Global Gravitational Stability}

Gas disks that are globally gravitationally
unstable develop bar mode perturbations which can rapidly transport angular momentum and
deliver large gas masses to the gravitational centers.  This drives 
rapid transformation of galaxy morphology, and as a result a compact massive stellar
system, a nuclear star cluster or a stellar bulge, develops in the
center.  \citet{Pawlik:11,Pawlik:13} found this process to be particularly
efficient in protogalactic disks assembling in $\sim (10^8-10^9)\,M_\odot$ halos,
assuming no reionization by external sources.  Objects forming in the reionized universe and
having reduced baryon fractions, however,
may avoid this fate. Indeed, only some of the most
luminous dwarf spheroidals in the Local Group, the
Sagittarius Dwarf and NGC 205, contain 
stellar nuclei; the others seem to be described by single-component
surface brightness profiles.  We proceed to asses the potential for bar mode
instability in the dwarf spheroidal progenitor objects.

For analytical simplicity of the forthcoming analysis, we
assume that $\lambda
/\sqrt{2} \ll
c^{-1}$ so that $R_{\rm disk}\ll r_{\rm s}$.   The
dark matter mass enclosed within radius $r\lesssim R_{\rm disk}$ is
$M_{\rm NFW} (r)\approx (400\pi/3)\rho_{\rm crit} c^3 r^2 r_{\rm s} /{\cal
  M}(c)$, where
$\rho_{\rm crit}$ is the critical density of the universe and ${\cal M}(x)=\ln(1+x)-x/(1+x)$.
The radial component of the gravitational force in the disk is
baryon dominated at radii where $\beta\int_0^r 2\pi \Sigma(R) R dR >
M_{\rm NFW}(r)$, where $\beta\approx \sqrt{r/R_{\rm disk}} \lesssim 1$ is a
reduction in the radial gravitational
potential of the disk due to its flattened nature.\footnote{The square root is
a fitting function approximating the value of $\beta$ in the specific
case of a razor-thin exponential disk embedded in a $\rho\propto
r^{-1}$ halo.}  This condition is most likely to be satisfied at
$r\approx 0.8\,R_{\rm disk}$.  At this reference radius, the condition becomes
\beq
\label{eq:baryon_domination}
\lambda<
\sqrt{\frac{0.5\,f_{\rm disk} f_{\rm b} {\cal M}(c)}{c^2}} ~ ~ ~ ~ ~ ~
(\textrm{baryon domination}) .
\eeq
For reference values $f_{\rm disk}=0.5$, $c= 4$, and a baryon fraction
substantially reduced from the cosmic mean $f_{\rm b}=0.1\,\Omega_{\rm b}/\Omega_{\rm
  m}$ (see Section \ref{sec:baryon_fraction}), baryon domination requires an improbably small disk spin
parameter, $\lambda\lesssim 0.01$.
For the typical disk
spin parameter $\lambda\sim 0.05$, dark matter gravity
dominates the radial gravitational force throughout unless the baryon
fraction is close to the cosmic mean.

In disks that are not geometrically thin, a refinement of the
criterion in Equation (\ref{eq:baryon_domination}) may be required.  If the rotationally supported disk is
thin and the radial pressure gradient can be neglected, the
disk is
stable to bar mode
excitation if the baryons dominate the radial component of the
gravitational force (e.g.,
\citealt{Christodoulou:95}, their Section 3.4.4).  If the gas pressure
gradient contributes significantly to the radial force balance, this
reduces the threshold for stability.  \citet{Fridman:84}, in their
Part V, Section 4.5.2, find that in a toy model of a uniformly
rotating disk embedded in a halo, the disk-to-halo mass ratio $M_{\rm disk}/M_{\rm
  halo}$ at which the instability sets in increases with the factor
$(1-\Pi)/(1-2\Pi)$, where $\Pi$ denotes the fractional reduction of the
gravitational force by the pressure force.  In this simplified model,
pressure stabilizes the bar mode for $\Pi\geq1/2$.   
With the vertical structure derived in Section
\ref{sec:vertical_structure}, the pressure averaged over one scale
height of the disk
is $P\sim \Sigma c_{\rm eff}^2/(2h)$ and the
pressure gradient acceleration is $a_P \sim c_{\rm eff}^2/R_{\rm disk}$.
For disks on the verge of global instability, we can compare \emph{twice} the
gravitational acceleration due to the halo, $a_{\rm grav} \sim
2GM_{\rm NFW}(r)/r^2$,
to the pressure gradient acceleration, to find $\Pi \sim a_P / a_{\rm
  grav}\sim1.4 (c_{\rm eff}/V_{200})^2
/\lambda_{0.05}$, where in the last step we substituted $c=4$ and $\lambda_{0.05}=\lambda/0.05$. This
confirms that in low-mass halos with low $V_{200}$, the global (bar
mode) instability can be partially stabilized by the radial pressure
force.\footnote{Consistent with this analysis, dwarf irregular
  galaxies, seen here as dwarf spheroidal satellites' analogs in the field, 
exhibit increasing relative disk thickness with decreasing
  luminosity \citep{Roychowdhury:13}.}

\subsubsection{Local Gravitational Stability}

Local stability of the H\,I disk requires $Q=\kappa c_{\rm eff}/(\pi G\Sigma) > 1$,
where $\kappa$ is the epicyclic frequency.  In
a disk of a finite thickness, and considering short wavelength radial perturbations
with wave numbers $k$, the effective gravity is reduced by the factor
$\sim (1+kh)^{-1}$, implying local stability for $Q>Q_{\rm
  crit}\sim 0.5-1$.  For rotational profiles due to the dark matter halo only,
$\kappa\sim \sqrt{3GM_{\rm NFW}(R)/R^3} \sim\sqrt{3}\Omega\propto R^{-1/2}$.  
Thus, if $c_{\rm eff}$ is approximately independent of radius and
$R_{\rm disk}\ll r_{\rm s}$, the disk is
the most susceptible to local instability at $R\sim R_{\rm disk}/2$;
for $R_{\rm disk}\sim r_{\rm s}$, the radius of highest
susceptibility shifts to $R\sim R_{\rm disk}$. 
In Figure \ref{fig:disk}, we plot $Q$ evaluated at $R=R_{\rm
  disk}/2$ in the dwarf spheroidal progenitor disks.  We find that
$Q_{\rm min}\gg Q_{\rm crit}$ at all parameter values.  This shows that due to 
the sharply suppressed baryon fractions, H\,I disks in the
common mass scale progenitor objects are stable to the
local gravitational instability of the disk.  The disks become
marginally unstable only
when the common mass scale parameter is at least twice the
S08 value, $M_{300}\gtrsim 2\times10^7\,M_\odot$.
Very speculatively, the local gravitational instability of the
H\,I disk, which starts setting in halos experiencing tidal
truncation of their mass assembly at masses $\sim
5\times10^9\,M_\odot$, may be associated with the development of
conditions for massive stellar cluster formation in the disk. Some of them may be the progenitors of the metal poor globular
clusters as in  the scenario of \citet{Kravtsov:05}.  The massive
clusters can also migrate into the galaxy center and build a nuclear
star cluster \citep[e.g.,][]{Agarwal:11}.

\subsubsection{Disk Edges}
\label{sec:edges}

The H\,I disk extends to the maximum radius at which the gas surface
density is sufficient for a neutral layer to be present that is
shielding itself from the intergalactic ionizing background.
Letting $f= 10^5\,f_5\,\textrm{cm}^{-2}\,\textrm{s}^{-1}$
denote the ionizing photon number per unit area per unit time and assuming that the disk midplane remains neutral and
opaque to ionizing photons, the number of photons absorbed per unit
area of the disk per unit time is $\sim f/4$.  The maximum disk surface
density that can be in the ionized state on each face of the disk is
given by the Str\"omgren condition $\alpha_{\rm rec}
[\frac{1}{2}\Sigma_{\rm HII} X / (h_{\rm HII} m_p)]^2 \sim \frac{1}{4}f /
h_{\rm HII}$.   Here,
$X\sim 0.75$ is the hydrogen mass fraction and $\alpha_{\rm rec}$ is
the recombination coefficient, which we take to be the average of the
case A and B coefficients $\alpha_{\rm rec}\sim 3\times10^{-13}\,\textrm{cm}^3\,\textrm{s}^{-1}$.  Denoting the effective sound speed in
the ionized gas with $c_{\rm HII}=10^6\,c_{\rm
  HII,6}\,\textrm{km}\,\textrm{s}^{-1}$ and the vertical scale
height of the ionized gas layer with $h_{\rm HII}\sim c_{\rm
  HII}/\mu_{\rm halo}$, we find that the critical
surface density for the presence of a neutral layer is $\Sigma_{\rm
  HII,crit} \sim (0.1-0.2)\,f_5^{1/2}\,c_{\rm
  HII,6}^{1/2}\,M_\odot\,\textrm{pc}^{-2}$, with the numerical
coefficient increasing with $M_{{\rm TT},\epsilon}$. This estimate assumes that the radius of the
  edge is much smaller than the halo scale radius $r_{\rm
    s}$. Relaxing this assumption would result in a lower $\mu_{\rm
    halo}$ and higher $\Sigma_{\rm HII,crit}$.
The critical surface density at which H\,I disk edges in more
massive systems are found
in the local universe, $\sim 0.4\,M_\odot\,\textrm{pc}^{-2}$, lies
slightly above this range.  The ionizing background  first increases
to $z\sim 2$ and then decreases toward even high redshifts
\citep[e.g.,][]{FaucherGiguere:09,Haardt:12}.\footnote{\citet{Schaye:04}
argues for a high effective value $f_5\sim10$ of the ionizing photon flux
in the local galaxies, but this seems high and possibly in
tension with the upper limit of \citet{Adams:11}.}
This could raise the critical disk edge surface density up to a maximum
of $\sim 1\,M_\odot\,\textrm{pc}^{-2}$.  Comparing these thresholds to
the surface densities in our model, we can conclude that 
H\,I disk edges in dwarf spheroidal progenitor objects will occur at $\sim(2-3)\,R_{\rm disk}$. We also
conclude that the disks in $M_{300}\lesssim 5\times10^6\,M_\odot$
halos with substantially reduced baryon fractions will not contain any
neutral gas over their entire history.

\subsubsection{Dependence on Angular Momentum}

The results of this section are potentially 
sensitive to the angular momentum of the gas that has settled in the
protogalactic disk.  For disk angular momenta exceeding the fiducial
value $\lambda=0.05$ assumed here, we expect the qualitative
conclusions to
remain unchanged, with a higher fraction of disk gas having surface
densities below the threshold for the WNM-to-CNM transition.  For atypically small angular momenta, $\lambda\ll
0.05$, however, the central column density could become high enough to
allow the atomic-to-molecular transition in the inner part of the disk,
potentially triggering intense central star formation, and potentially
even explosive baryon removal by SNe (see Section \ref{sec:simmering} below).   The randomness
inherent in hierarchical merging guarantees that $\lambda$ samples
the full range of values, occasionally dipping
well below our fiducial choice, especially in certain major mergers
between halos.  While
disk buildup from
cold-mode accretion in more massive halos produces characteristically
high values of spin parameter $\lambda\sim 0.1$ \citep{Stewart:13},
dedicated cosmological simulations are required to assess the angular
momentum content of the baryons in small, reionized halos

\subsection{From Tidal Truncation to Ram Pressure Stripping}

\citet{Mayer:06} have determined that dwarf spheroidal satellites with maximum
circular velocities $V_{\rm max}<30\,\textrm{km}\,\textrm{s}^{-1}$ are
completely ram pressure stripped if their orbits have pericenters of
$\lesssim 50\,\textrm{kpc}$ from the center of the Milky Way.
Generally, as we have remarked in Section \ref{sec:common_mass_scale},
the redshift at which the gas is ram pressure stripped
will be somewhat lower than that of the tidal truncation of the halo's
mass assembly $z_{\rm
  TT}(M_{\rm TT})$. In the interval $z_{\rm ram}<z<z_{\rm TT}$,
the neutral gas in the halo may temporarily remain, allowing
some final star formation to proceed.  If tidal truncation takes place
when the common mass scale object has approached to
within $r\sim 3\,r_{200,{\rm MW}}$ of a Milky Way progenitor halo of radius
$r_{\rm 200,{\rm MW}}$, and the ram pressure stripping happens when the object
has approached to within $r\sim r_{200,{\rm MW}}$ (formally, the point of
incorporation into the Milky Way's or Andromeda's satellite population), 
then we can estimate the time an
object infalling radially from rest at a large distance 
traverses this radial range.  We assume
that the outer NFW profile approximates the
spherically averaged mass distribution to multiple
virial radii \citep[see, e.g.,][]{Masaki:12}.  
The time
interval has a negligible dependence on the halo concentration 
and approximately
equals 
\beq
\Delta t_{{\rm TT}\rightarrow{\rm ram}} \sim 0.13\,H(z)^{-1} ,
\eeq
 where
$H(z)$ is the Hubble parameter at redshift $z$.  This time is
of the same order of magnitude as the growth time of the infalling
halo prior to tidal truncation. Thus, if
other processes, such as ram pressure stripping by large-scale
structure flows and galactic outflows, do not remove the star forming
gas before the infall,
star formation in the
period between tidal and ram pressure stripping should be taken into
account when estimating the final stellar mass of the
object.  

\subsection{Simmering Star Formation and the Unlikeliness of
  Outflows}
\label{sec:simmering}

The local gravitational stability of the H\,I disk does not
imply the absence of star formation, because the WNM, if overpressured by turbulence, can become thermally unstable
and cool by metallic fine structure line emission to form
transient concentrations of cold neutral gas.  In pressure
equilibrium, the low-filling-factor CNM is $\sim100$ times denser than
the WNM in which it is entrained and has a Jeans
length shorter by the same factor \citep[e.g.,][]{Vazquez-Semadeni:09,Vazquez-Semadeni:12,Saury:13}.  Stars can form when turbulence
collects a sufficient mass of CNM to cross the threshold for
gravitational instability in the CNM alone.  The typical Jeans masses
are of the order of $M_{\rm J}\sim10^3\,M_\odot$, which is marginally sufficient for the
collapsing gas clump, upon turbulent gravitational fragmentation, to form stars sampling the entire stellar initial
mass function (IMF).  However, because of the stochasticity of the local CNM buildup,
larger star forming complexes are unlikely to form, consistent with
the observation that dwarf galaxies in the local universe (typically
more massive than the common mass scale progenitor systems) seem to form only
a small fraction of their
 stars in the starburst mode \citep{Lee:09a}.
Given this, we expect that
the star formation proceeded very slowly in typical dwarf spheroidal progenitor disks, with at
most a few isolated $\sim M_{\rm J}$ sites, each containing a handful of massive
young stars, being present at any time.  
The feedback from such low-grade, distributed star
formation in the form of H\,II regions and SNe seems 
insufficient to drive explosive removal of baryons from the halos.
At best, it can remove gas from the disk and deposit it in
   the enveloping diffuse gaseous halo, from where it can condense back into the
   disk, enriching it with the nucleosynthetic
   return of preceding stellar generations, now diluted through
   turbulent mixing in
   the halo (see Section \ref{sec:chemical_evolution} below).
Therefore, \emph{we ignore the feedback and assume that the dwarf spheroidal
progenitor objects do not lose their baryons to outflows}, in contrast
with the outflows 
detected in, say, the substantially more massive Ly$\alpha$ emitting and Lyman break
galaxies.  

The assumption of negligible feedback may seem to contradict some
 analyses suggesting that SNe efficiently drive baryons from atomically
cooling halos, as it may be naively expected on purely energetic grounds.  A closer
consideration, however, reveals that clumpiness of the protogalactic gas
diminishes the destructive impact of the SNe, in particular if
they explode with energies similar to those of the typical
core-collapse SNe in the nearby universe, rather than the
ultra-energetic and much more destructive pair-instability SNe
(see, e.g., \citealt{Greif:07,Wise:08}).  For example, \citet{Wise:12}
carried out a high resolution cosmological simulation tracking
star formation, reionization, SNe, and chemical enrichment in
halos during the epoch of reionization. Their representative 
``quiet'' halo attaining a final mass of $\sim
10^8\,M_\odot$ at $z=7$ still manages to
retain a baryon fraction $f_{\rm b}\sim 0.5\,\Omega_{\rm
  b}/\Omega_{\rm m}$, in spite of having converted
$\gtrsim10^5\,M_\odot$ of its baryons
into stars; their ``intense'' $\sim10^9\,M_\odot$ halo retains close
to all of its baryons while forming an $\sim2\times10^6\,M_\odot$ star
cluster.  The dwarf spheroidal progenitor halos we
envision have reduced baryon fractions at the outset.
Therefore, we expect that their star formation would have proceeded even more
quietly and, thus, that the impact of SNe on the baryon budget
would have been even weaker.

Dwarf irregular galaxies residing in more massive dark matter halos than those of the 
Milky Way's dwarf spheroidal satellites do exhibit starbursts and coherent
galactic outflow signatures \citep{Martin:98,Martin:99}.  These outflows can be
interpreted as having been triggered by periodic destabilization and central
channeling of gas in mergers.  We
expect that in these
more massive dwarf galaxies with higher $V_{\rm max}$,  baryon
fractions are closer to the cosmic mean and the gas disks are closer to the threshold
for global gravitational instability that, if triggered by a merger
event, can produce a concentrated starburst.

\section{Results}
\label{sec:results}

\subsection{Stellar Masses and Half-Light Radii}

Assuming that neutral gas turns into stars on the time-scale
$\tau_{\rm dep}$ given by Equation (\ref{eq:depletion_time}), the
star formation rate is
\beq
\label{eq:M_dot_star}
\dot{M}_\star = \frac{M_{\rm HI}}{\tau_{\rm dep}} ,
\eeq
where $M_{\rm HI}$ is the H\,I mass in the most massive
progenitor branch of a halo's merger tree.  We ignore any star formation that may
have occurred in the minor branches of the merger tree representing
smaller halos merging with the main branch.  This
approximation is justified by the strong dependence in Equation
(\ref{eq:f_b_Gnedin}) of the baryon fraction at any
given redshift on the mass of a low-mass halo.  The minor halos
will on average have severely suppressed baryon fractions and will be
unable to sustain self-shielding H\,I disks and form stars of
their own.  We also approximately neglect the actual depletion of the neutral
gas to star formation.  Given the long depletion time, this is
clearly a good approximation for objects ending star formation at $z\gg1$, but
even at lower redshifts, the depletion of the H\,I disk is
compensated in part by the stellar mass return and recombination from
the fraction $(1-f_{\rm disk})$ of the baryonic content that remains
outside the disk.  Therefore
\beq
\label{eq:HI_mass}
M_{\rm HI} \approx f_{\rm disk}\,f_{\rm b} \,M_{\rm halo} ,
\eeq
and with this, the zero-age main sequence (ZAMS) stellar mass of a
dwarf spheroidal is 
\beq
M_{\star} = M_{\star,{\rm TT}} + M_{\star,{\rm TT}\rightarrow{\rm
      ram}} ,
\eeq
where
\bea
\label{eq:stellar_mass_integral}
M_{\star,{\rm TT}} &\approx& \frac{f_{\rm disk}}{\tau_{\rm dep}} \int_{z_{\rm TT}}^\infty
f_{\rm b} (M_{\rm MMP}[M_{\rm TT},z_{\rm TT};z],z) \nonumber\\ &
&\times\, M_{\rm MMP}[M_{\rm TT},z_{\rm TT};z]\,
\frac{dt}{dz} \,dz 
\eea
is the stellar mass produced until the point of the tidal truncation of
the object's mass assembly, $dt/dz=(1+z)^{-1}H(z)^{-1}$, and
\bea
M_{\star,{\rm TT}\rightarrow{\rm
      ram}} \approx \frac{f_{\rm disk}}{\tau_{\rm
    dep}} f_{\rm b} (M_{\rm TT},z_{\rm
 TT}) \, M_{\rm TT}\,\Delta
t_{{\rm TT}\rightarrow{\rm ram}} 
\eea
is the star formation taking place 
after tidal truncation but before the halo's gas has been
ram pressure stripped.

\begin{figure}
\begin{center}
\includegraphics[width=0.475\textwidth]{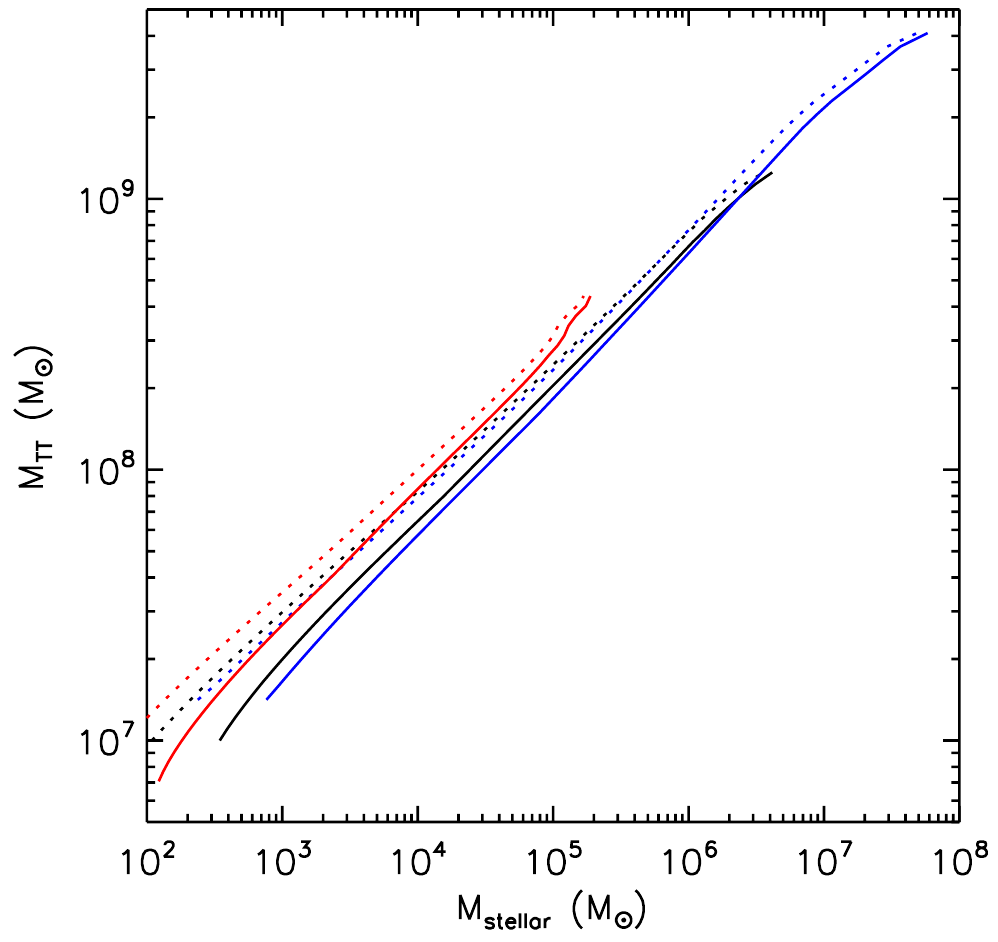}
\caption{The halo mass at the point at which an external tidal field
  truncated the mass assembly of the dwarf spheroidal progenitor halo as a function of the stellar mass for
  halos on the common mass scale relation with
  $M_{300}=(2^{-0.5},\,2^0,\,2^{+0.5})\times10^7\,M_\odot$ (red,
  black, and blue lines, respectively), excluding and including star formation in
  the period between tidal and ram pressure stripping (dotted and
  solid lines, respectively). }
\label{fig:mass_mass}
\end{center}
\end{figure}

Figure \ref{fig:mass_mass} shows the dependence of $M_{{\rm
  TT},\epsilon}$ on $M_\star$ for three representative common mass
scale families, $\epsilon =(-0.15,\,0,\,+0.15)$, we assume that
a factor of $f_{\rm ZAMS}=0.5$ of the ZAMS stellar mass remains at $z=0$
after stellar mass loss.  The $M_{\rm
  TT,\epsilon}(M_\star)$ is largely independent of the central density
parameter $\epsilon$.
A variation of $\approx 2\,\textrm{dex}$ in halo mass
  corresponds to a much larger variation of $\approx
  4\,\textrm{dex}$ in stellar mass, so that, approximately
\beq
M_\star
\label{eq:stellar_halo_prop}
  \propto M_{\rm TT}^2 . 
\eeq
This can be understood as arising from
  the scaling of the stellar mass with \emph{both the mass of the host
    halo, and the duration of the time available for
  star formation}, where the latter is itself a roughly linearly increasing
  function of the  halo mass.  Including the parametric dependence on
  the halo mass fraction in the disk $f_{\rm disk}$ and the gas-to-star conversion time-scale $\tau_{\rm dep}$, we have
\bea
\label{eq:stellar_halo}
M_\star &\approx& 2\times10^4\,M_\odot\,\left(\frac{f_{\rm
      disk}}{0.5}\right)\,\left(\frac{f_{\rm
      ZAMS}}{0.5}\right)\,\nonumber\\ & & 
\times\left(\frac{\tau_{\rm dep}}{10\,\textrm{Gyr}}\right)^{-1}\left(\frac{M_{\rm
      TT}}{10^8\,M_\odot}\right)^2 ,
\eea 
again approximately independent of the central density quantified by
$M_{300}$.\footnote{The stellar mass-halo mass relations in Equations
  (\ref{eq:stellar_halo_prop}) and (\ref{eq:stellar_halo_prop}) are
  significantly shallower than the ad hoc relations ranging in slope from $M_\star\propto
  M_{\rm TT}^{2.5}$ to $M_\star\propto M_{\rm TT}^{3}$ adopted
  elsewhere \citep{Koposov:09,Kravtsov:10,Ocvirk:11,Rashkov:12} to explain the
  properties of the dwarf population.  The ostensible success of the steep scaling
  in reproducing the dwarf luminosity function is consistent with the
  conclusions of \citet{Brook:13} that straightforward halo
  mass-stellar mass 
  abundance matching mandates a steep scaling $M_\star\propto M_{\rm
    TT}^{3.1}$. However the difficulty with reproducing the observed
  halo densities and maximum circular velocities
  \citep{BoylanKolchin:11,BoylanKolchin:12} suggests that 
  the assumptions of completeness and monotonicity entering the
  straightforward abundance matching may be questionable.}

\begin{figure}
\begin{center}
\includegraphics[width=0.475\textwidth]{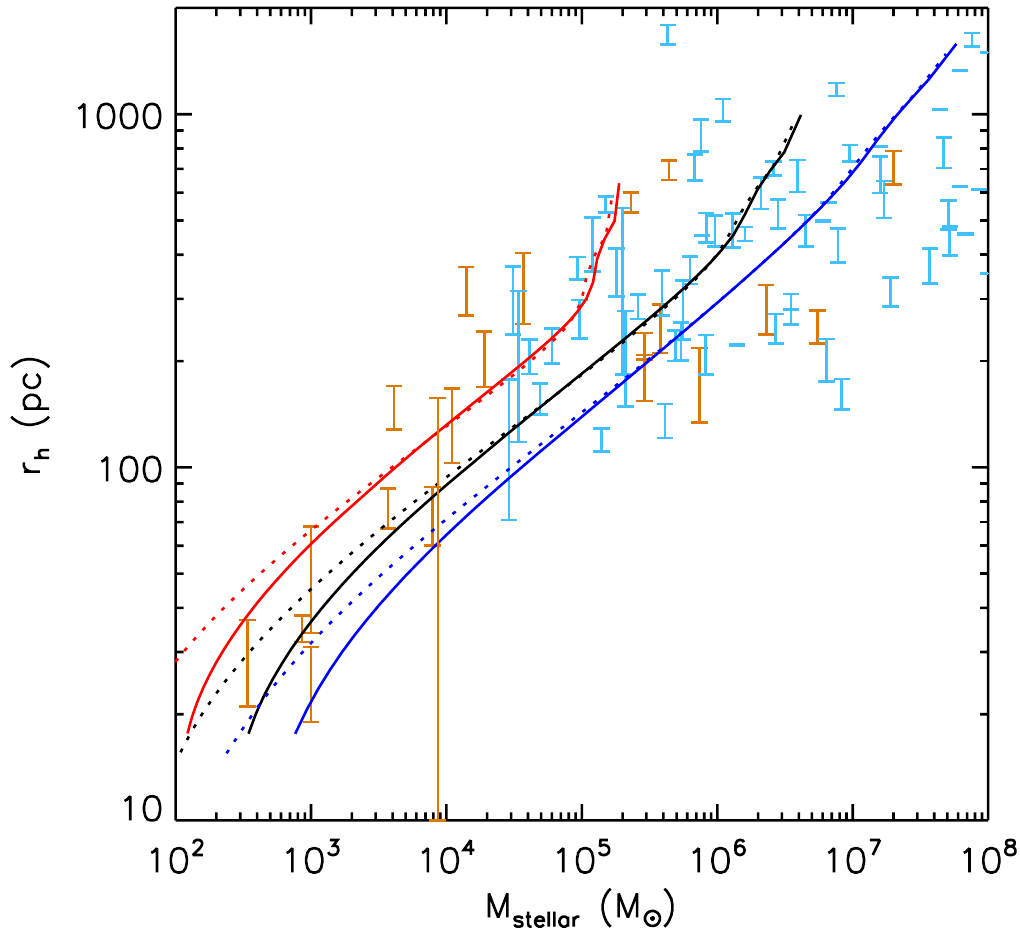}
\caption{The half-light radius as a function of the stellar mass for
  halos on the common mass scale relation with
  $M_{300}=(2^{-0.5},\,2^0,\,2^{+0.5})\times10^7\,M_\odot$ (red,
  black, and blue lines, respectively), excluding and including star formation in
  the period between tidal and ram pressure stripping (dotted and
  solid lines, respectively).  The data points denote dwarf spheroidal
 satellites of the Milky Way (brown) and Andromeda (light blue) from
 \citet{McConnachie:12}.  The half-light radius uncertainties were
 arbitrarily assigned to zero where McConnachie does not quote
 uncertainties.  The stellar masses assume solar mass-to-light ratios $M_\star/L=M_\odot/L_\odot$; the actual mass-to-light ratios could
 be up to a factor of $\sim 2$ higher.}
\label{fig:mass_radius}
\end{center}
\end{figure}

Figure \ref{fig:mass_radius} plots our calculated dwarf
spheroidal half-light radii as a function of the present-day stellar
mass. The figure shows the half-light radii and
stellar masses of the dwarf spheroidal satellites of the Milky
Way and Andromeda in the compilation of \citet{McConnachie:12}.  
We assume that the half-light radius is
related to the exponential scalelength of the disk via
$r_{\rm h}=1.7\,R_{\rm disk}$ where $R_{\rm
  disk}=0.05\,r_{200}/\sqrt{2}$, and arrive at the final value of
half-light radius by computing the stellar mass-weighted average of
$r_{\rm h}$ over each most massive progenitor track.
We also assume that the stellar mass-to-light ratio has
the solar value, $M_\star/L = M_\odot/L_\odot$.  The latter two assumptions are almost certainly not
very accurate. The stellar IMF will likely evolve \citep[see][]{Geha:13} as the
 metallicities of the gas disks increase with time and the CMB
 temperature floor drops.  In our crude model, the dependence on the
 IMF is encapsulated in the parameter $f_{\rm ZAMS}$, with
 more top-heavy IMFs implying lower $f_{\rm ZAMS}$ and $M_\star$. 
Theoretical work is in progress to develop
 a theory of the stellar IMF at low column densities and metallicities
 and high redshifts \citep[][]{SafranekShrader:14}. 

The  stellar masses and half-light radii provided by the model, especially
 allowing for a relatively small, $0.15-0.3\,\textrm{dex}$ variation of the common mass scale
 parameter $M_{300}$ around the
 S08 value,
 seem to be in close agreement with those of the observed dwarf spheroidal satellites.  This reproduces the observed $\approx
 5\,\textrm{dex}$ variation of the stellar mass across the dwarf
 spheroidal satellite sequence in the Milky Way.  The ultra-faint dwarfs reside in dark
 matter halos that were near, or slightly above, the atomic cooling
 threshold, 
\beq
M_{\rm halo,UFD}\lesssim2\times
 10^8\,M_\odot  ,
\eeq
when they formed the bulk of their stars.

\subsection{Chemical Evolution}
\label{sec:chemical_evolution}

We argued in Section \ref{sec:simmering} above that star formation in
the dwarf spheroidal progenitor objects takes place in the simmering
rather than the starburst mode, and that such slow star formation is
not likely to drive outflows powerful enough to remove the star
formation's nucleosynthetic return from the halo.  This hypothesis has immediate
implications for the objects' chemical evolution.  To explore these implications,
we estimate the dependence of the metallicity on the stellar mass.
Let
$f_{\rm ret}$ denote the fraction of the stellar mass returned by
winds and SNe, and let $Z_{\rm ret}$ be the metallicity of the
returned mass.  Assuming a negligible contribution from an initial (Pop III) pre-enrichment
that would have not left behind surviving low-mass stars, as
well as
instantaneous return (i.e., through
core-collapse ejecta and other potential prompt mechanisms) 
and instantaneous and homogeneous mixing in the entire gas mass of the
halo, we have that the metallicity of the gas in the halo is given by 
\beq
\label{eq:Z_from_Z_return}
Z_{\rm gas}\sim Z_{\rm ret}\, \frac{f_{\rm ret} \, M_\star}{f_{\rm b}
  \,M_{\rm halo}} ,
\eeq
where $M_{\star}$ and $f_{\rm
  b}$ are the stellar mass and the baryon fraction 
in the halo of mass $M_{\rm halo}$.  It is important to note that Equation
(\ref{eq:Z_from_Z_return}) allows for baryons to be added by
minor branches of the merger tree and by accretion from the IGM, but
it assumes that the added baryons are in the form of a chemically
pristine gas 
and that this gas mixes instantaneously with
the metal-enriched gas residing in the main branch. 

The average stellar metallicity will
then be given by
\bea
Z_{\star,{\rm TT}} &\sim& \frac{1}{M_{\star,{\rm
      TT}}}\int_{z_{\rm TT}}^\infty Z_{\rm gas,MMP}(z')\nonumber\\ & & \times
\dot{M}_{\star,{\rm MMP}} (z')\, \frac{dt}{dz'} \,dz' .
\eea
Substituting Equations (\ref{eq:M_dot_star}) and (\ref{eq:HI_mass}) we obtain
\beq
\label{eq:Z_star}
Z_{\star,{\rm TT}} \sim f_{\rm disk}\, f_{\rm ret} \,Z_{\rm ret}
\frac{\tau_{\rm rep,TT}}
{\tau_{\rm dep}} ,
\eeq
where
\beq
\label{eq:tau_star_tidal}
\tau_{\rm rep,TT}= \frac{1}{M_{\star,{\rm TT}}}\int_{z_{\rm TT}}^\infty   M_{\star,{\rm MMP}}(z')  \,\frac{dt}{dz'}\,dz' 
\eeq
is a ``reprocessing time,'' a measure of the time-scale on which the
object has become enriched with metals.  The reprocessing in the
period from the onset of tidal truncation to the final ram-pressure
stripping of the gas can be included by modifying Equation
(\ref{eq:Z_star}) to read
\beq
\label{eq:Z_star_CMS}
Z_{\star} \sim f_{\rm disk}\, f_{\rm ret} \,Z_{\rm ret}
\frac{\tau_{\rm rep}}
{\tau_{\rm dep}} ,
\eeq
where
\bea
\label{eq:tau_star_CMS}
\tau_{\rm rep} &=& \frac{1}{M_{\star}}\bigg[ ( \tau_{\rm rep,TT} +\Delta t_{{\rm
    TT}\rightarrow{\rm ram}} ) M_{\star,{\rm TT}} \nonumber \\ &
& + \frac{1}{2} (\Delta t_{{\rm
    TT}\rightarrow{\rm ram}})^2 \dot M_{\star}(z_{\rm TT})\bigg] .
\eea

\begin{figure}
\begin{center}
\includegraphics[width=0.475\textwidth]{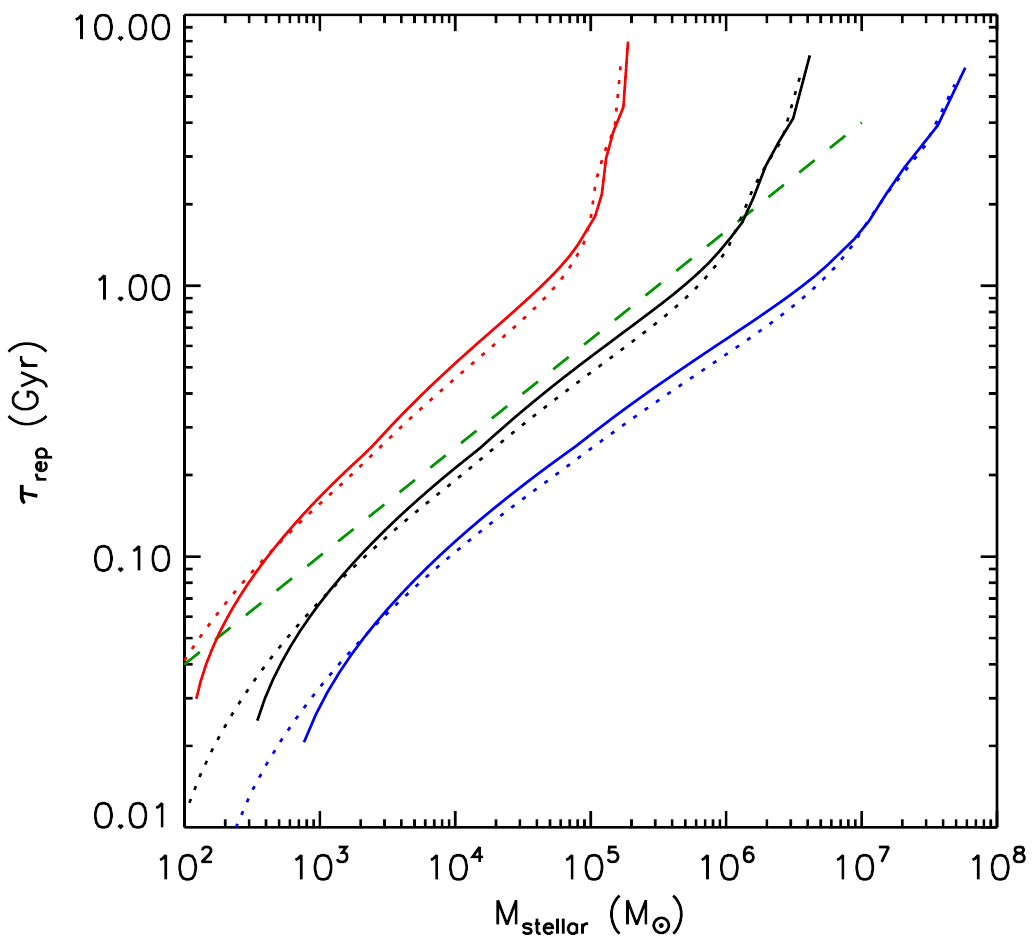}
\caption{The self-enrichment time-scale $\tau_{\rm rep}$ as a function of the stellar mass for
  halos on the common mass scale relations with
  $M_{300}=(2^{-0.5},\,2^0,\,2^{+0.5})\times10^7\,M_\odot$ (red,
  black, and blue lines, respectively), excluding and including star formation in
  the period between tidal and ram pressure stripping (dotted and
  solid lines, respectively).  The green dashed line of arbitrary
  normalization indicates the
  slope of the $M_\star$ -- $\tau_{\rm rep}$ relation required for
  Equation (\ref{eq:Z_star}) to reproduce the iron abundance-stellar mass
  correlation $Z_{\rm Fe}\propto M_\star^{0.4}$ in the Milky Way dwarf
  spheroidals of \citet{Kirby:11a}
  under the artificial assumption that the iron synthesis rate is
  proportional to the instantaneous star formation rate. }
\label{fig:tau_stellar}
\end{center}
\end{figure}

The metallicity $Z_\star$ in Equation (\ref{eq:Z_star_CMS}) is proportional to a product
of several uncertain factors and cannot be predicted robustly. 
Therefore, we refrain from directly plotting $Z_\star$. Instead, in Figure
\ref{fig:tau_stellar} we plot the arguably more robust reprocessing times $\tau_{\rm
  rep,TT}$ and $\tau_{\rm rep}$ as functions of the stellar
mass for three neighboring common
mass scale families. The times vary from mere tens of Myr at the 
low-mass end to almost $10\,\textrm{Gyr}$ at the high-mass end.  The
reprocessing times as a function of the stellar mass exhibit a sharp
upturn when the time exceeds $\sim 1\,\textrm{Gyr}$, corresponding to
halos with $z_{\rm TT}\lesssim 2$.  The upturn mass increases
from $10^5\,M_\odot$ to $10^7\,M_\odot$ for $M_{300}$ increasing from
$0.7\times10^7\,M_\odot$ to $1.4\times10^7\,M_\odot$.

For illustration, with the specific choices $f_{\rm disk}=f_{\rm ret}=0.5$ and
  $Z_{\rm ret}=0.05$ (all of which are uncertain), we have 
\beq
Z_{\star} \approx \frac{\tau_{\rm rep}}{\tau_{\rm dep}}\,Z_\odot ,
\eeq
implying that the metallicities of the dwarf spheroidals
resulting from the prompt, core-collapse-type enrichment
should vary over two orders of magnitude, from $\sim
10^{-2.3}\,Z_\odot$ (for $M_{300}=10^7\,M_\odot$ and $M_\star\gtrsim 10^3\,M_\odot$) to
just below $Z_\odot$.
This degree of metallicity
 variation seems similar to that found in the observed
  stellar mass-metallicity relation \citep{Kirby:11a}, where the dwarf
  iron abundance scales with the stellar mass as $Z_{\rm Fe} \propto
  M_\star^{0.4}$.  The stellar mass dependence of the reprocessing
  time in Figure \ref{fig:tau_stellar} seems to agree with this slope
  (green dashed line in the figure) up to the point of
  upturn.  If the average iron yield of core-collapse SNe is
  $Z_{\rm ret,Fe}/Z_{\rm ret}\sim 0.1$ and the solar iron abundance is
  $Z_{\odot,{\rm Fe}}/Z_\odot\approx 0.2$, then the model predicts an
  average iron abundance spanning the range
  $-2.6\lesssim \langle [{\rm Fe}/{\rm H}] \rangle< -0.3$.  

Allowing
  for SNe Ia to introduce an iron abundance enhancement of $\Delta \langle [{\rm
    Fe}/{\rm H}] \rangle \sim +0.3$ at the high-mass end of the Milky dwarf
  spheroidal sequence, but not at the low-mass end (since these dwarfs
  are ram pressure stripped on time-scales relatively short compared
  to the typical SN Ia delay times), the range widens to match the observations
  \citep[see][]{Kirby:11a}.
  The statistics of delay times $\tau_{\rm Ia}$ for the
  onset of SN Ia enrichment is currently poorly determined and thus it
  is not possible to carry out a systematic comparison with the
  reprocessing times $\tau_{\rm rep}$.
  All dwarf spheroidals seem to exhibit signatures of SN Ia enrichment
  except for the ultra-faint dwarfs Segue 1 and  Ursa Major II, which
  exhibit $\alpha$-enhancement in all the stars with medium-resolution
  spectroscopy-based abundance measurements, consistent with pure core-collapse nucleosynthesis
  \citep{Vargas:13}.  If the delay time is, say, $\tau_{\rm Ia}\sim 0.1\,\textrm{Gyr}$,
then our model indeed allows for pure core-collapse enrichment only in
the
faintest dwarfs, consistent with the observations
\citep[see, also,][who reached a similar, slightly less restrictive
conclusion based on the
analysis of a smaller stellar sample]{Frebel:12}.

We do not compute the metallicity distribution functions and are not
in the position to compare the predictions of this model with the
logarithmic and linear stellar metallicity scatter of the observed
dwarfs.  The assumption of instantaneous mixing implied in Equation
(\ref{eq:Z_from_Z_return}) precludes a realistic computation of the
scatter.  A more accurate approach, which we defer to further study,
would include a model in which chemical enrichment is stochastic and
non-instantaneous \citep[e.g.,][]{Oey:00,Oey:03,Pan:07}. Finally, we
note that because of the metal dilution in the gas mass accumulating until the onset
of tidal truncation, it seems unlikely that potential
pre-enrichment by Pop III SNe would have significantly affected the
mean metallicity of the galaxy, but it could have certainly left
its imprints in the structure of the low-metallicity tail of the
metallicity distribution.

\section{Where Are the Densest Satellites?}
\label{sec:true_fossils}

\citet{BoylanKolchin:12}  compared the densities
of the Milky Way's dwarf spheroidal satellites derived from stellar
kinematical measurements to those of the
most massive satellites of Milky Way-equivalent galaxies in a
cosmological simulation, and found that the simulated halos contained satellites that were denser than the observed dwarfs.
The simulated halos contained at least $10$ subhalos with
maximum circular velocities $V_{\rm max}>
25\,\textrm{km}\,\textrm{s}^{-1}$, higher than allowed for objects
lying on the common mass scale relation (see Section
\ref{sec:common_mass_scale} and Figure \ref{fig:tidal}).  These objects have been called ``too
big to fail'' \citep{BoylanKolchin:11}, because of having deeper
gravitational potential wells 
than the observed common mass
scale objects, which are consistent with $V_{\rm max}\sim
(10-30)\,\textrm{km}\,\textrm{s}^{-1}$ \citep{Strigari:10};
they should have retained even higher baryon fractions
and formed more luminous, easily detectable stellar systems \citep[e.g.,][]{Kravtsov:04}. 

There have been attempts to resolve the too-big-to-fail problem by
considering the possibility that the dwarf spheroidal satellite population of the
Milky Way is a statistical outlier or that it reflects an overestimate
of the mass of the Milky Way's dark matter halo.
\citet{Purcell:12} find that a subsample of realizations of Milky Way-analog
halos in CDM simulations have satellite densities consistent with the observations, and argue
that the problem with densities can be explained on statistical
grounds.  \citet{Strigari:12}, however, find that the Milky Way is not
a statistical outlier in its number of bright satellites as compared
to similar galaxies in {\it Sloan Digital Sky Survey}.  Others
\citep[e.g.,][]{Wang:12,VeraCiro:13} attempt to address the problem by
invoking the possibility that the mass of the Milky Way halo is 
lower than normally assumed, which reduces the expected number of satellites with high
maximum circular velocities.

We propose a different, very speculative solution to the too-big-to-fail
problem, that a number of satellite subhalos with circular velocities
above the range consistent with the common mass scale objects are
indeed present in the Local Group, but that the stellar systems in these
``too-big-to-fail'' satellites
are not being identified with the dwarf spheroidal morphological
type, but with an altogether different type of stellar system.  Recall
that because the most massive progenitor
accretion histories of halos with similar central densities are
themselves similar (Section \ref{sec:universal_history}), the most
massive progenitors of the subhalos
with $M_{300}\gg 10^7\,M_\odot$ will have already had masses $M_{\rm MMP}\gtrsim
10^8\,M_\odot$ at $z=10$, and as such, they will have been able to
form their first stellar generations before reionization and to retain baryon
fractions $\sim \Omega_{\rm b}/\Omega_{\rm m}$ after
reionization. 
It is our general expectation that in gas-rich
early halos retaining high
baryon fractions, especially those with gas accretion times much shorter than the
typical gas-to-stars conversion time at high surface densities $\sim
1\,\textrm{Gyr}$, 
global gravitational instabilities facilitate rapid
angular momentum transport and drive large gas masses into the halo
centers.  This is clearly seen in the simulations of \citet{Pawlik:11,Pawlik:13} tracking the
formation of a $10^9\,M_\odot$ halo at $z=10$ with no external
ionizing sources,
where the bar mode instability transported between a quarter and a
third of the baryons in the halo into the inner few
tens of parsecs.

The morphological type of the resulting
stellar system will differ from that of a non-nucleated dwarf
spheroidal galaxy.  The system will at least contain a dense central stellar
cluster.  The star formation that produced the system will have been much more intense
than in the extended, gravitationally stable disk
in baryon-poor halos described in Section
\ref{sec:star_formation_in_dwarfs}.  This intense nuclear star
formation might drive an explosive removal of baryons from the halo,
possibly foiling star formation altogether outside the central cluster.
The Local Group already contains dense,
centrally concentrated stellar systems including the compact
elliptical galaxy M32 and the massive
globular clusters $\omega$ Centauri of the Milky Way and Mayall II
(or G1) of Andromeda.
It also contains less dense, spheroidal stellar systems 
with embedded dense nuclear stellar 
components, such as the
nucleus of the spheroidal galaxy NGC 205 and the nuclear globular cluster
M54 of the Sagittarius dwarf spheroidal galaxy. 
There is currently no evidence for dark matter in
$\omega$ Cen and G1, but a dark matter halo with a density comparable to that of the densest
dwarf spheroidal galaxies is consistent with the kinematic data 
(K.~Gebhardt, priv.~comm.). 

Interestingly, the stellar metallicity spreads in $\omega$ Cen and M54
are larger than those in typical 
globular clusters and are similar to the spreads
in dwarf galaxies \citep[e.g.,][]{Leaman:12,Willman:12}; 
the same may be
true for G1 \citep{Meylan:01}.  The large
spreads can be interpreted as indirect
evidence that the stars in these clusters formed in
the gravitationally confining central density cusps of dark matter halos, rather than, say, in
the fragmentation of locally gravitationally unstable extended galactic gas disks, or
in the collision of gas streams in galaxy mergers, the latter two being the
standard mechanisms thought to produce globular clusters.  

It is often
taken for granted that to reconcile a galaxy-like origin with the
compact present form of stellar systems like $\omega$ Cen, 
an outer, more extended, dwarf spheroidal-like, low-surface-density stellar
component had to have been in place and to have subsequently been stripped from the
dense nuclear star cluster.  However, if the low-surface-density spheroidal
components are products of star formation in a reionized universe, and
the nuclear component is a pre-reionization fossil, then it seems
possible that the nuclear cluster-like dense stellar systems could have
formed without galaxies surrounding them.

\section{Conclusions}
\label{sec:conclusions}

We have constructed an analytical model describing star
formation in the progenitors of dwarf spheroidal satellite galaxies in the
Local Group.  The model combines input from published simulations of
halo mass assembly in the $\Lambda$CDM universe
with a star formation prescription consistent with the results
of investigations of star formation in gas-rich, low-metallicity dwarf
galaxies in the local universe.  Our main conclusions are as follows.

Parametrizing the dynamical mass profile of dwarf spheroidal
satellite galaxies with $M_{300}$, the mass enclosed within the
innermost $300\,\textrm{pc}$, and with 
the help of the halo concentration dependence on the mass and
redshift calibrated by \citet{Prada:12}, 
we derived families of mass-redshift pairs on which the
dwarf spheroidals' host
halos must have lied at the critical time at which an external tidal
field truncated their mass assembly.

Computing the mean most massive progenitor histories for dark-matter-dominated galaxies
with the same $M_{300}$, thus ostensibly belonging in a common mass scale
family, we find that they are similar in the
sense that the mean most massive progenitor mass at a fixed redshift
$z$ varies very little
among objects with very different masses at the lower redshifts at
which their mass assembly was truncated by an external tidal field.  This reflects an early assembly of
the central $300\,\textrm{pc}$ of the host halos.

The mean most massive progenitor histories of the objects with central
densities currently low enough that $M_{300}\lesssim 10^7\,M_\odot$
became able to form stars
at lower redshifts than those at which the Local Group is expected to have
undergone reionization.  Therefore, we conclude that the dwarfs formed
most of their stars under reionized conditions.  The objects with only
somewhat higher central densities
$M_{300}\gtrsim 2\times10^7\,M_\odot$, on the other hand, almost
certainly formed their first stellar populations before reionization.
This led us to hypothesize that it is the temporal relation to reionization
that defines the character of a galaxy, whether it will at $z=0$ be
recognized as a dwarf spheroidal (forming post-reionization), or a
different, still to be determined morphological type (forming its
first stellar populations before reionization).

Informed by recent numerical investigations of the evolution of 
halo baryon fractions in a patch of the universe that has experienced
reionization, we investigated the dependence of the baryon fraction on
 the central density parameter $M_{300}$ and the halo mass and the
 point at which an external tidal field truncated the mass
assembly of these halos.  We found that halos with the same $M_{300}$ have
similar baryon fractions independent of the halo mass,
but that the baryon fraction is very sensitive to $M_{300}$, dropping
by an order of magnitude for a factor of 2 decrease in $M_{300}$.

The sensitivity of the baryon fraction to the central density led us
to suggest
an explanation for why dwarf spheroidal satellites of the Milky
Way fall on the approximate common mass scale relation
$M_{300}\approx 10^7\,M_\odot$.  The satellite halos with only
somewhat lower central densities, 
$M_{300}\lesssim 0.5\times 10^7\,M_\odot$, had baryon fractions too low
for a self-shielding H\,I of a sufficient surface density to have
been present in the halo prior to the ram pressure stripping.  On the
other hand, the satellite halos with only somewhat higher central
densities, $M_{300}\gtrsim 2\times 10^7\,M_\odot$, commenced efficient
star formation before reionization and after the reionization was
complete,
retained baryon
fractions near the cosmic mean. The high baryon fractions meant that
the gas in these galaxies was dense enough to
form giant molecular clouds, and that it was globally
gravitationally unstable. The bar mode instability
transported substantial gas masses into the very centers of the latter
halos,
where it formed stellar systems much more compact than the dwarf
spheroidals, but potentially resembling massive globular clusters or the
nuclear star clusters in spheroidal galaxies.

Having argued that ionized and warm neutral gas flows in the
dwarf spheroidal progenitor objects were both globally and locally
gravitationally stable, and that they were partially rotationally
supported with disk-like morphologies, 
we assessed the conditions affecting star formation in these flows (surface
densities, metallicities, UV backgrounds)
and suggested that the formation of the dwarf spheroidals' stars
resembled the star formation currently taking place in the
lowest-baryonic-mass dwarf galaxies in the local universe and near the
outer edges of late-type disk galaxies.  Specifically, the stars
formed in the predominantly atomic phase of interstellar medium with a
characteristic gas depletion time of $\sim 10\,\textrm{Gyr}$.
Metallic fine structure line emission, rather than molecular emission,
facilitated the cooling of the gas. We also argued that the star formation
proceeded in small units, implying that the resulting feedback should not have had a particularly destructive effect
on the gas content of these halos, in contrast with published analyses
suggesting intense outflows from dwarf spheroidals' progenitors.

By integrating star formation rates over mean most massive progenitor
histories, we computed the stellar masses and half-light radii
expected to be found in satellite halos belonging in
the common mass scale family $M_{300}=10^7\,M_\odot$ and
representative neighboring families.  We found that the stellar mass was
a steep, approximately quadratic
function of the mass of the progenitor halo at the point of
the tidal truncation of mass assembly, just prior to its incorporation
into the substructure of the more massive host galaxy.  The steepness
is a consequence of the strong, approximately linear dependence (at a
fixed $M_{300}$) of the
time available for star formation on the maximum, tidally truncated mass
of the halo.  

Allowing for a small $|\Delta \log M_{300}|\lesssim 0.3$ variation of the
central density (and an even smaller variation $|\Delta \log
M_{300}|\lesssim 0.15$ at the low luminosity end), 
the stellar masses and half-light radii determined from 
our crude model agree with
those observed among the dwarf spheroidals in the Milky Way and Andromeda.
The success of the crude model suggests that the ultra-faint dwarfs
have
exceptionally low stellar masses and high mass-to-light ratios 
because they formed in relatively
low mass, ``atomically cooling'' halos ($\sim 10^8\,M_\odot$) with
baryon fractions reduced to $\sim10$ per cent of the cosmic mean. The star
formation in these objects lasted
only $\lesssim 100\,\textrm{Myr}$ before the gas was ram pressure stripped.

Pursuing our hypothesis that the chemical enrichment in the
dwarf spheroidals' progenitors operated in the lossless regime,
we computed the cumulative enrichment due to the prompt (e.g., core collapse)
nucleosynthetic sources.  The resulting metallicities exhibit similar
magnitudes and a similar
scaling with the stellar mass as the observed dwarfs.

\section*{Acknowledgments}

We would like to thank Mia Bovill, Matt McQuinn, and Louis Strigari
for useful comments and insightful conversations that have
helped us improve the manuscript.  We would also like to thank 
Francesco Prada and Miguel S{\'a}nchez-Conde for clarifications
on dark matter halo concentrations, Andrei
Mesinger, Emanuele Sobacchi, and Takashi Okamoto for clarifications on
halo baryon fractions, and Alan McConnachie for advice on interpreting
kinematic mass measurements.  We also acknowledge helpful
conversations with Michael Boylan-Kolchin, James Bullock, Karl
Gebhardt, Anatoly
Klypin, John Kormendy, Chalence Safranek-Shrader, Risa Wechsler, and Beth Willman. 
This research was supported by NASA through Astrophysics Theory and Fundamental Physics Program grant NNX09AJ33G and through NSF grant AST-1009928.

\footnotesize{

}

\end{document}